\documentclass[review]{elsarticle}
\usepackage{geometry}
\usepackage{graphicx}
\usepackage[utf8]{inputenc}
\usepackage{amsmath}
\usepackage{amssymb}
\usepackage{geometry}
\usepackage{textcomp}
\usepackage{xcolor}
\usepackage{slashed}
\usepackage{dcolumn}
\usepackage{bm}
\usepackage{titlecaps}
\usepackage{xcolor}
\usepackage[colorlinks,citecolor=blue]{hyperref}
\usepackage{multicol}
\usepackage{multirow}
\usepackage{array}
\usepackage{url}

\newcolumntype{C}[1]{>{\centering\let\newline\\\arraybackslash\hspace{0pt}}m{#1}}
\newcolumntype{M}[1]{>{\centering\arraybackslash}m{#1}}

\begin{document}

\begin{frontmatter}

\title{Measurements of gamma ray, cosmic muon and residual neutron background fluxes for rare event search experiments at an underground laboratory}
\author{Sayan Ghosh\corref{corrauthor}}
\cortext[corrauthor]{Corresponding author}
\ead{sayan.ghosh@saha.ac.in}

\author{Shubham Dutta}
\ead{shubham.dutta@saha.ac.in}

\author{Naba Kumar Mondal}
\ead{naba.mondal@saha.ac.in}

\author{Satyajit Saha}
\ead{satyajit.saha@saha.ac.in}

\address{Saha Institute of Nuclear Physics - HBNI, I/AF Bidhan Nagar, Kolkata 700064, India.}

\begin{abstract}
Ambient radiation background contributed by the penetrating cosmic ray particles and the radionuclides present in the rock materials have been measured at an underground laboratory located inside a mine at 555 m depth. The laboratory is being set up to explore rare event search processes, such as direct dark matter search, neutrinoless double beta decay, axion search, supernova neutrino detection, etc., that require specific knowledge of the nature and extent of the radiation environment in order to assess the sensitivity reach and also to plan for its reduction for the targeted experiment. The gamma ray background, which is mostly contributed by the primordial radionuclides and their decay chain products, have been measured inside the laboratory and found to be dominated by rock radioactivity for $E_\gamma \lesssim 3 \,{\rm MeV}$.  Shielding of these residual gamma rays for the experiment was also evaluated. The cosmic muon flux, measured inside the laboratory using large area plastic scintillator telescope, was found to be: $(2.051 \pm 0.142 \pm 0.009) \times 10^{-7}\, {\rm cm}^{-2}.{\rm sec}^{-1}$, which agrees reasonably well with simulation results. The neutron background flux has been measured for the radiogenic neutrons and found to be: $(1.61 \pm 0.03) \times 10^{-4} \, {\rm cm}^{-2}.{\rm sec}^{-1}$ for no threshold cut. Detailed GEANT4 simulation for the radiogenic neutrons and the cosmogenic neutrons have been carried out.  Effects of multiple scattering of both the types of neutrons within the surrounding rock and the cavern walls were studied and the results for the radiogenic neutrons are found to be in reasonable agreement with experimental results. Neutron fluxes contributed by those neutrons of cosmogenic origin have been reported as function of the energy threshold. 
\end{abstract}


\end{frontmatter}


\section{Introduction}
\label{intro}
Rare event search experiments, such as direct dark matter search, neutrinoless double beta decay, axion search, supernova neutrino detection, etc. look for extremely small and often unknown interactions\cite{ref1,ref2,ref3,ref4}. An essential requirement for establishing such experiments is to assess the ambient radiation background caused by various naturally occuring radioactive sources specific to the chosen site, termed as the radiogenic background, and also that caused by the penetrating cosmic ray particles and the secondary radiation caused by them, known as the cosmogenic background. While the radiogenic background depends on the concentration of primordial radionuclides (mostly $^{40}$K, $^{238}$U and $^{232}$Th) and their decay chain products present at the local rock surrounding the underground laboratory cavern, the cosmogenic background has direct correlation with the depth of the laboratory, physical properties such as density of the rock, morphology of the rock and to some extent, on the topography of the terrain above ground. Therefore, measurements of the radiogenic background is essential before establishing an underground laboratory for rare event search at any depth. On the other hand, precise knowledge about the nature, directionality and energy spectra of the cosmic ray particles reaching the Earth is readily available\cite{rpp2017,GeisHonda} and passage through the layers of rock up to the cavern can be traced in detail using tracking and simulation tools (eg. GEANT4\cite{Geant4Pap}). Therefore, the cosmogenic background can be assessed in great detail using the cosmic ray flux data, particle tracking and simulation. However, the role of actual measurements cannot be underestimated\cite{mks}.       

Among all the fundamental and the light composite particles, gamma rays, electrons, alpha particles and neutrons pose the most common and often irreducible background which affect the outcome of highly sensitive experiments of importance in physics. The irreducible background is caused by the intrinsic radioactive impurity inside the active detector medium or the associated materials in the proximity, which are generally kept inside the shielding enclosure meant for mitigating the reducible radiation background. Therefore, it is essential to know and quantitatively assess all such radiation background before setting up the experiment. For the direct search for dark matter experiments looking for signals from very tiny elastic and inelastic recoil of nuclei, the background neutrons are the most worrisome since they also cause nuclear recoil of similar kinetic energies\cite{RevwRef,Mei_Hime}. Alpha particles, produced from the intrinsic impurities due to the radionuclides present in the detector materials also cause undesired background. For neutrinoless double beta decay (NDBD), the gamma ray background, especially near the $Q_{\beta\beta}$-value region is the most important one, but the neutron background, especially the cosmogenic neutron background can also be problematic since they can cause cosmogenic activation of interfering nuclei\cite{ref6a}. Neutrino experiments, particularly the ones that measure the neutral current interactions, such as the SNO experiment\cite{ref7} are also sensitive to the neutron background, apart from the irreducible background caused by the alpha particles, electrons and gamma rays.             

Since the nature of the reducible radiation background is specific to the site and related to the type of experiments to be set up at the site, measurements of various types of radiation background have been done and compared with detailed Monte Carlo simulation to understand the systematics of various types of radiation at the Jaduguda Underground Science Laboratory (JUSL), located at a depth having vertical rock overburden of 555 m (1604 metre water equivalent) in Jharkhand, India. The laboratory, essentially a prototype laboratory for rare event search type experiments, is a cavern of approximately 4.5 m length, 4.5 m width, and 2.2 m average height. In the simulation model, the laboratory is approximated as a hemispherical cavern of 2.2 m radius.  The gamma ray background due to the content of predominantly $^{238}$U, $^{232}$Th and $^{40}$K nuclei in the surrounding rock was measured inside the cavern. The corresponding gamma ray shielding requirements may be optimized based on the outcome of these measurements.    
Cosmic muons from the high energy tail of the atmospheric muon spectrum reach inside the laboratory and create additional neutron as well as gamma ray background. The residual cosmic muon flux was measured inside the laboratory using large area plastic scintillator telescope and corroborated the results through detailed Monte Carlo simulation. The cosmogenic neutron flux and their spectral distribution, were evaluated using Monte Carlo simulation with input from the above experimental results. In situ measurements of the radiation background at the underground laboratory, their comparison with those measurements carried out at the overground laboratory, interpreting the results and understanding their systematics through detailed Monte Carlo simulations are reported in this paper. 	

\section{Gamma ray background measurements}
\label{sec:Gamma}
The gamma ray background was measured with a Thalium activated Cesium Iodide (CsI(Tl)) detector of 50 mm diameter and 50 mm height, made available from the Crystal Technology Section of Bhabha Atomic Research Centre (BARC), Mumbai\cite{CsI}. The scintillator crystal was optically coupled directly to Hamamatsu R1306 photomultiplier (PMT) with built-in ASIC board to provide bias voltage to the PMT and a multichannel analyzer (MCA). The built-in electronics was directly addressable through USB port of a laptop or a PC running the MCA software. The detector system was made portable so that it can be carried inside the underground facility to all the available depths and operated for at least 2 hours without plugging in to any power source. The temperature of the detector assembly was also logged by the MCA to compensate for the shift in energy calibration. The detector was calibrated using standard gamma sources (such as $^{137}$Cs, $^{60}$Co, $^{22}$Na), and also $^{238}$U and $^{232}$Th enriched materials. The detector system was programmed to collect data in batches of 2 hours at a stretch for a few days. Data were collected for equal time period inside the underground laboratory and at the surface laboratory above the mine complex. 

\begin{figure}[ht]
	\centering
	\includegraphics[scale=0.5]{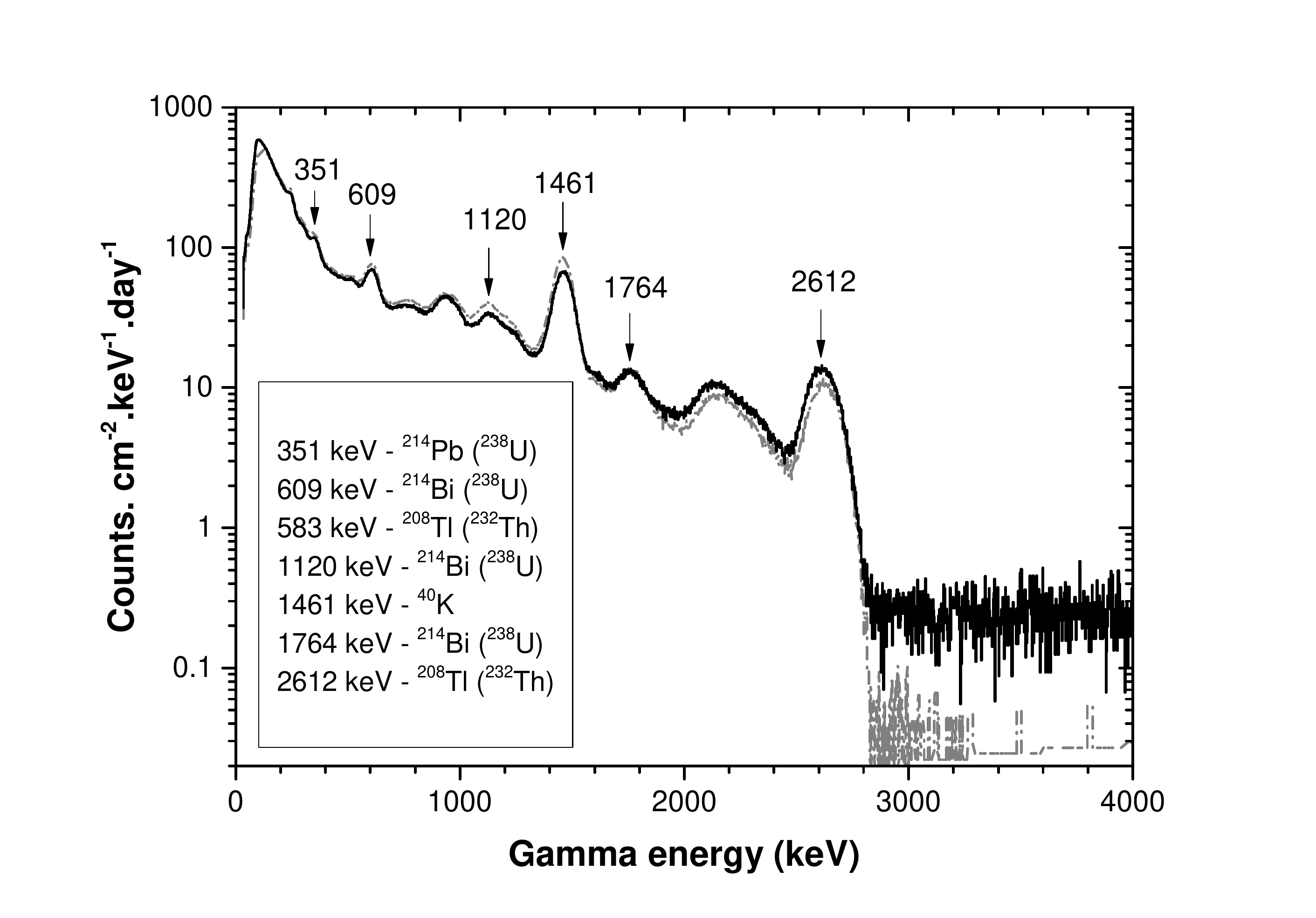}
	\caption{Gamma ray spectra measured using CsI(Tl) detector at the underground laboratory (grey line) and at the surface laboratory (black line).}
	\label{gam1}
\end{figure}
Gamma ray energy spectra, obtained at the two sites and corrected for detector efficiency and temperature variations, are shown in the Fig.~\ref{gam1}. Discrete gamma ray peaks or the photopeaks due to $^{40}$K, and decay chain products from the $^{238}$U and $^{232}$Th remnants at both the sites are found to have similar intensities but with subtle differences. The acquired spectra were segmented into 8 zones for evaluating the zone-wise shielding requirements to arrive at the optimum shield thickness. Each zone is 500~keV wide, except for the zones 1 and 2 which are 300 and 200 keV spans, and zone 8 which is of about 1600 keV span. The zone divisions are shown in Table~\ref{Tb1}. The ratio of intensities, defined as $S=I_u/I_s$, where $I_s$ is the intensity measured above ground and $I_u$ is the intensity measured at underground by the same detector, is termed as the suppression factor($S$), and were evaluated over each zone. It can be observed that the $^{40}$K peak at 1461 keV within zone 4 clearly indicates excess presence of $^{40}$K at the underground site as compared to the surface. Similar excess in intensity, albeit a bit less, was also seen in zone 3, which could be due to the gamma rays at 583- and 609-keV arising from the U/Th decay chain nuclei ($^{208}$Tl and $^{214}$Bi). However, the reduction in intensity at the underground site over the zone 7, which includes the photopeak at 2612-keV due to $^{208}$Tl, is apparently in contrast with what was observed in zone 2. It may be noted that the intensity had drastically reduced at the underground site to almost 2\% of that at the surface laboratory over the zone 8 ($E_\gamma \gtrsim 3\,{\rm MeV}$) and therefore, it is expected that the corresponding Compton background, which would add to the intensity over zone 7, would also be reduced. It may be mentioned here that the gamma ray background over zone 8 is mostly due to the continuum type bremsstrahlung radiation arising from the showers generated by the penetrating cosmic ray particles, which would give large contribution at the surface.     

\begin{table}[ht]
\begin{center} 
\begin{tabular}{llllcccc}
\hline \hline
Zone & Range                      & Count rate  & Suppression  & \multicolumn{4}{c}{Shielding factor at UG for the different} \\ 
        & of $E_\gamma$        & at UG        & factor at UG  & \multicolumn{4}{c}{Lead shield thickness (cm)} \\
& (MeV)             & ($\rm cm^{-2}.sec^{-1}$)  & ($S$) & 5  & 10 & 20 & 30 \\
\hline
\\
1 & $0-0.3$      &  0.9028(32) & 0.9202(46) & $\sim10^{-6}$ & $\sim10^{-7}$ & $<10^{-7}$ & $<10^{-7}$  \\   
2 & $0.31-0.5$ &  0.2146(15) & 1.0762(96) & 0.000614 & $1.921\times10^{-6}$ & $\sim10^{-7}$ & $<10^{-7}$ \\
3 & $0.51-1.0$ &  0.2796(18) & 1.0753(99) & 0.0201 & 0.00146 & $1.655\times10^{-5}$ & $1.428\times10^{-7}$ \\
4 & $1.01-1.5$ &  0.2177(16) & 1.176(13) & 0.0632 & 0.0101 & 0.000256 & $1.066\times10^{-5}$ \\
5 & $1.51-2.0$ &  0.06858(89) & 0.984(18) & 0.0970 & 0.0207 & 0.000888 & $5.493\times10^{-5}$ \\
6 & $2.01-2.5$ &  0.0333(6) & 0.804(20) & 0.122 & 0.0292 & 0.00157 & $9.753\times10^{-5}$ \\
7 & $2.51-3.0$ &  0.02095(49) & 0.754(23) & 0.143 & 0.0354 & 0.00251 & 0.000156  \\
8 & $3.01-4.6$ &  $1.06(35)\times10^{-4}$ & 0.0226(76) & 0.176 & 0.0435 & 0.00274 & 0.000194 \\
\hline \hline
\end{tabular}
\end{center}
\caption{Assessment of zone-wise gamma ray shielding at the underground (UG) laboratory due to different thicknesses of Lead. 
Count rates are given in counts. cm$^{-2}$.\,sec$^{-1}$. The suppression factor $S$ is defined in the text.}
\label{Tb1}
\end{table}

A typical rare event search experiment demands for passive shielding from such gamma ray background using high $Z$ elements, such as Lead. Since the attenuation of gamma rays of different energies in Lead vary with thickness, and the fact that the gamma rays typically undergo multiple Compton scattering and also the high energy gamma rays ($E_\gamma \gtrsim 1 {\rm MeV}$) cause pair production, a straightforward exponential mass attenuation formula would not apply in determining the effective thickness of the shielding materials. We have attempted a systematic study to evaluate the optimum thickness needed for shielding in the gamma radiation environment at the underground laboratory using GEANT4 tracking simulation as detailed below. 

A rectangular and hollow lead box was constructed as the shielding material surrounding a cylindrical Caesium Iodide (CsI) detector (20 cm diameter $\times$ 20 cm long). Gamma rays of specific energies according to the zone classification (see Table~\ref{Tb1}) were allowed to impinge from random directions on the outer walls of the Lead box. Only those $\gamma$-rays reaching the detector were recorded and counted. For each energy zone, $10^7$ events were generated by the event generator and the shielding factors were calculated as the ratio of the number of recorded events over the number of events initially generated. The shielding factors for different thicknesses of Lead for the various $\gamma$-ray energy zones are presented in Table \ref{Tb1}.

It can be seen from Table~\ref{Tb1} that for the gamma rays background flux up to the zone 7 ($E_\gamma \lesssim 3\, {\rm MeV}$) can be reduced to $\sim 10^{-6}\, {\rm counts.\,cm^{-2}.\,sec^{-1}}$ with 30 cm thick Lead shield. Largely the effects of the gamma rays background due to U--Th decay chain products and the primordial nuclei (such as $^{40}$K) can be drastically reduced by almost 6 orders of magnitude by the passive shield. Active shield using plastic scintillators in the veto mode can reduce it further.



\section{Cosmic Ray Muons}
\label{muons}
Muons, created from cosmic ray interaction in the upper atmosphere, are the most abundant cosmic ray charged particles at the Earth's surface, with a mean energy of $\sim$4 GeV and an integrated intensity of 1 $\rm cm^{-2}\hspace{1mm}min^{-1}$~\cite{PDG_Cosmic,Tanabashi:PDG}. In this energy regime, the muons behave as minimum ionizing particles and therefore, they are highly penetrating. They penetrate through large distances inside the Earth and give rise to radiation background, which interferes with many rare event search type of experiments set up at underground laboratory facilities. Moreover, the interaction of muons with the rocks around the underground laboratories produce neutrons
through muon spallation reactions\cite{spall} and muon induced fission\cite{mufiss}. These energetic neutrons produce nuclear recoils having energy comparable to those expected to be caused by the proposed dark matter candidates, popularly known as weakly interactiong massive particles (WIMPS).  Therefore, it becomes very important to know the flux and the energy spectrum of muons in order to predict the nature of cosmogenic neutron background.

\subsection{Experimental Setup and Results}
\label{expt}
A set of four fast plastic scintillators (Saint Gobains BC-404) was used for the muon flux measurement. Each scintillator, having dimensions of 96.4 cm $\times$ 32.1 cm $\times$ 1 cm, was coupled to a 50 mm diameter photomultiplier (ET Enterprises 9807B) using wavelength shifting (WLS) optical fibers running along the length of the upper surface\cite{neha}. 
\begin{figure}[ht]
	\centering
	\includegraphics[scale=0.7]{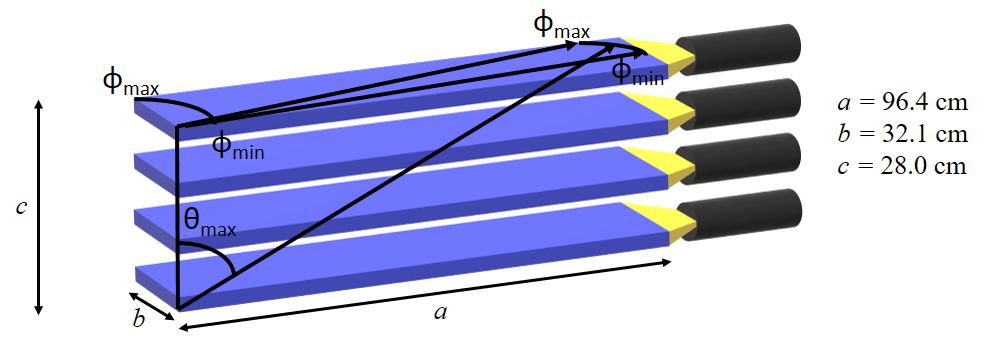}
	\caption{Schematic representation of the scintillation detector showing the co-ordinate system and geometric parameters used for numerical estimation.}
	\label{DetScheme}
\end{figure}
The scintillators were placed in a vertical stack, as shown in Fig. \ref{DetScheme}, with a gap of about 8 cm between two adjacent ones. The signal from each photo-multiplier was split into two parts, one was fed into a leading edge discriminator while the other was fed into a charge to digital converter or QDC in short, for charge integration. In the event of passage of a muon through all the scintillators, a four fold coincidence would be registered, provided each output signal passes through the preset discriminator threshold cut.To begin with, the detection efficiency of each scintillator of the stack was maximized with respect to the other three operated in coincidence, by tweaking the bias voltages of the corresponding photomultiplier and the signal discriminator threshold. This was done at the over ground laboratory. 
\begin{figure}[ht]
\centering
\begin{picture}(470,160)
\put(0,0){\includegraphics[scale=0.4]{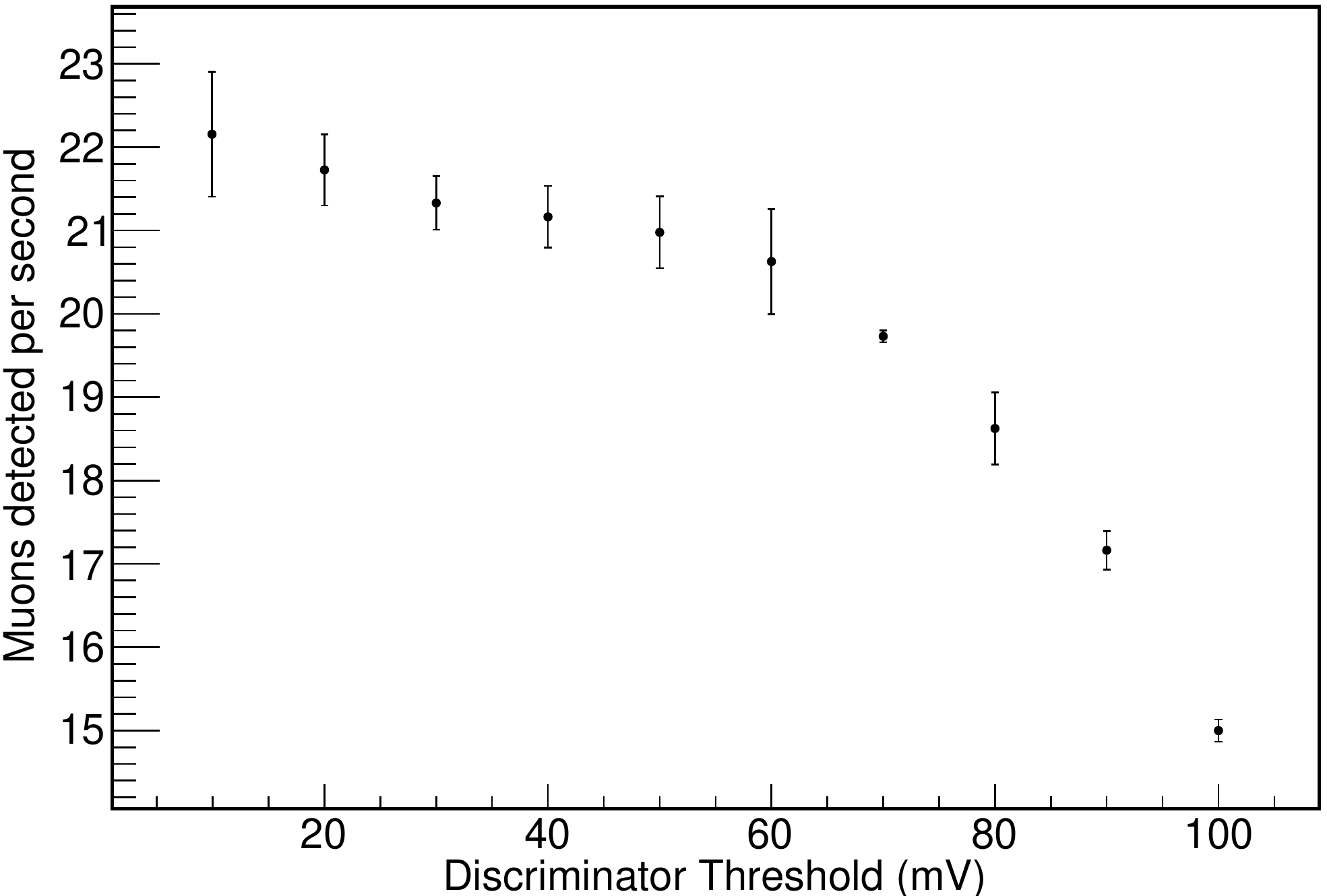}}
\put(40,50){\bf (a)}
\put(240,0){\includegraphics[scale=0.45]{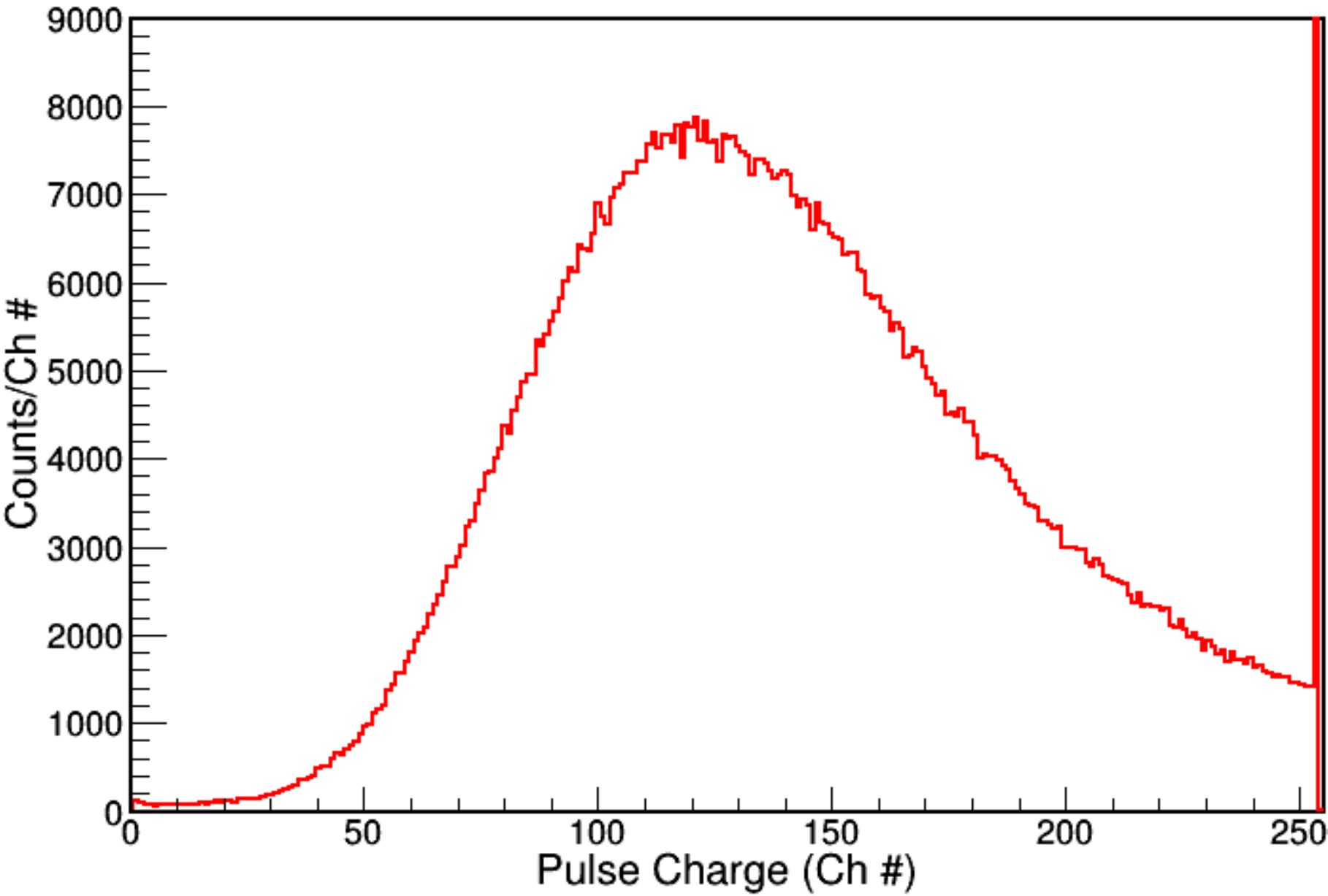}}
\put(280,50){\bf (b)}
\end{picture}
\caption{(a) Muon detection rate (4-fold coincidence) as a function of discriminator threshold. (b) The energy deposition spectrum of muons recorded by one of the scintillators when the detector stack was placed at the surface laboratory.}
\label{SurfOp}
\end{figure}
 
\begin{figure}[ht]
\centering
\includegraphics[scale=0.35]{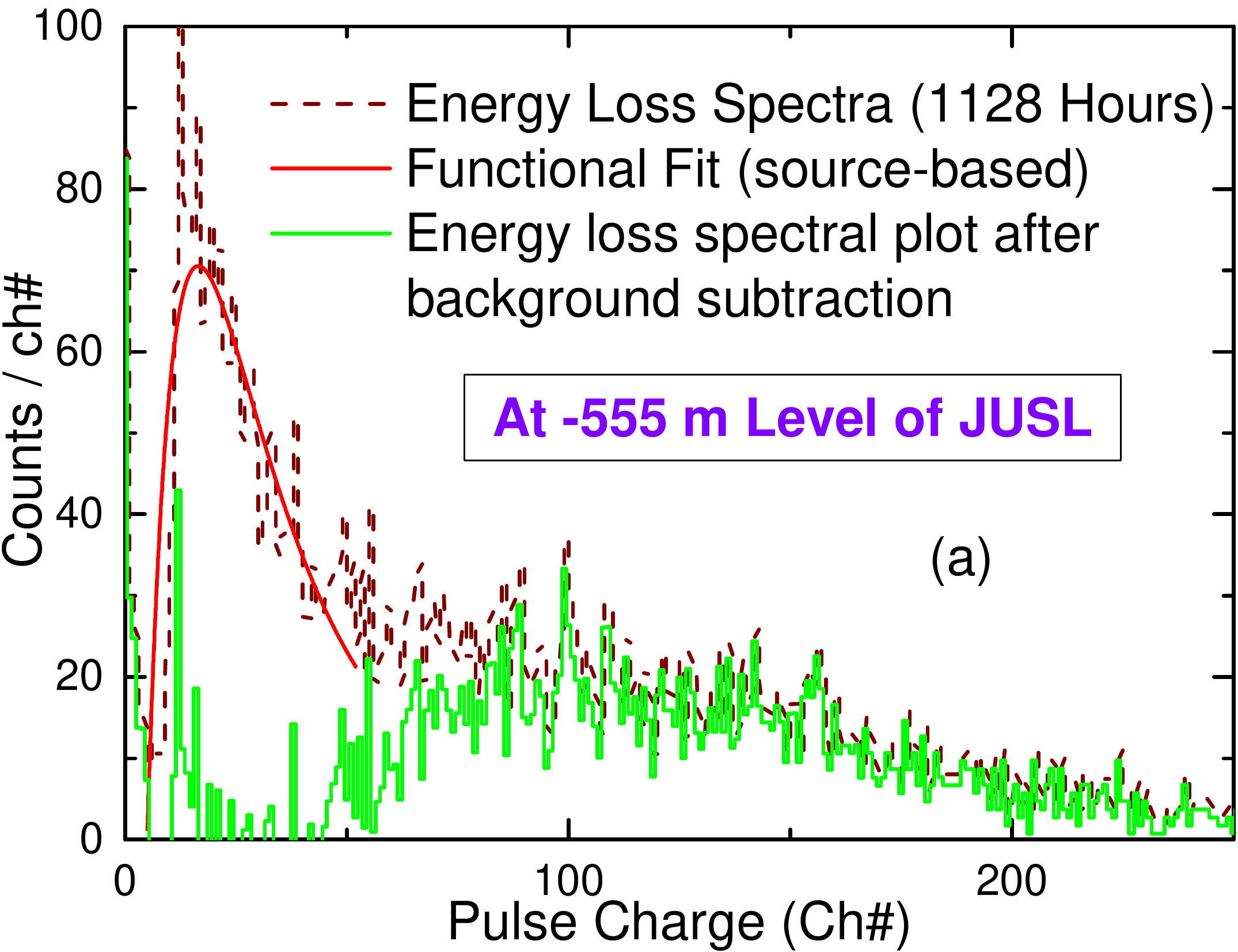}
\hspace{2 mm}
\includegraphics[scale=0.35]{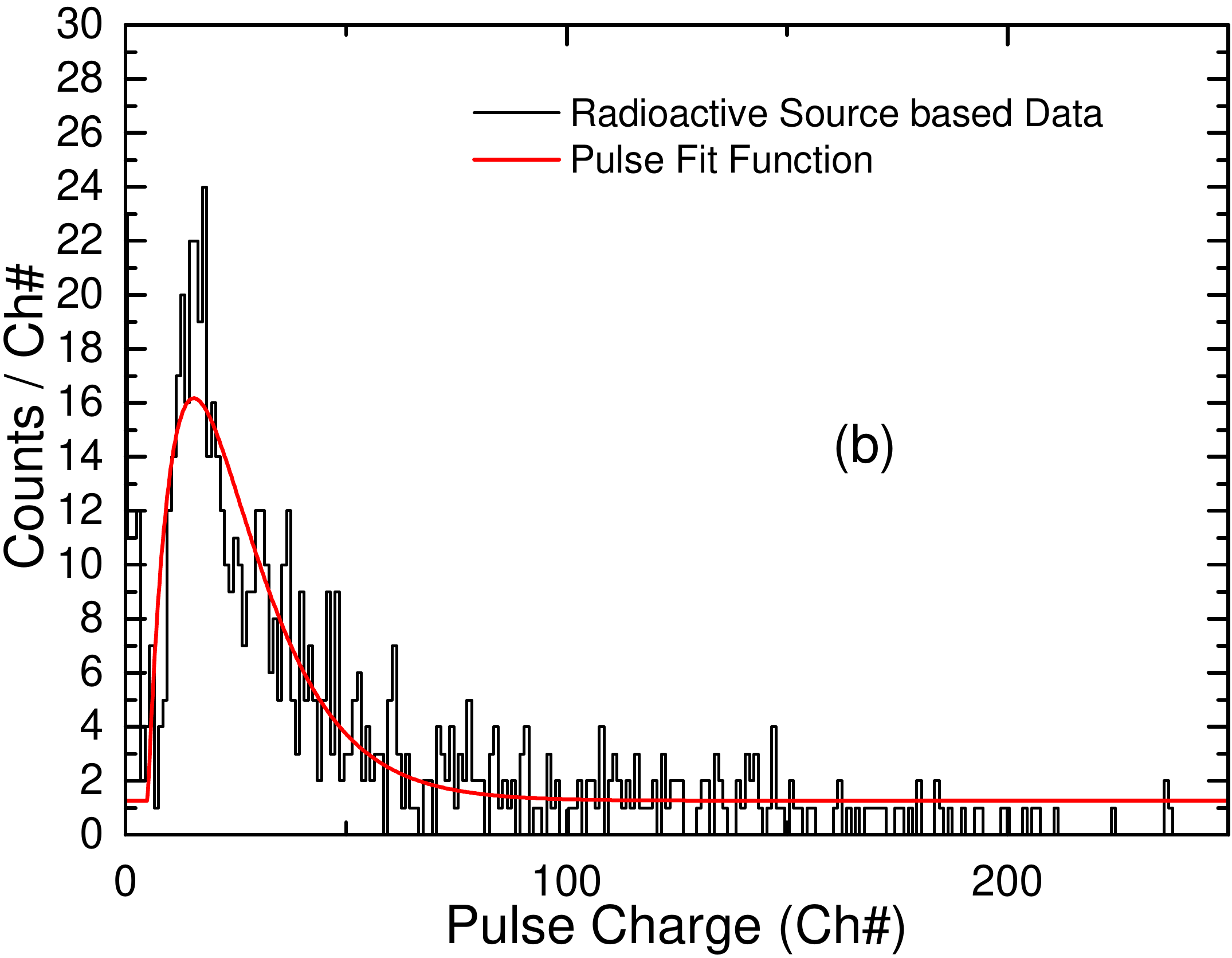}
\caption{(a) The muon energy deposition spectrum as measured in 555 m deep underground lab. The background subtracted spectrum is indicated inside the plot. (b) Energy deposition spectrum for a $\rm Na^{22}$ gamma source. }
\label{UGSpecExp}
\end{figure}
The overall efficiency of the scintillator stack in 4-fold coincidence mode, turns out to be $(97.04\pm1.17)$\%. Likewise, the common discrimination threshold for all 4 scintillators was set by raising it till the point where the detection rate starts to fall off (see Fig.~\ref{SurfOp}(a)). Subsequently, we have operated the full setup in the 4-fold coincidence mode to obtain the energy deposition ($\Delta E$) spectrum of the muons for all the scintillators. The $\Delta E$ spectrum is shown in the Fig.~\ref{SurfOp}(b) for one of the scintillators for five minutes of data acquisition. It can be seen qualitatively that the $\Delta E$ spectrum of muons follows a Landau nature, which is typical of the energy loss spectrum of cosmic charged particles in an absorber medium. 

We took the entire experimental setup to the underground site at 555 m (1604 mwe) depth to measure the muon flux inside the underground laboratory. The spectral distribution of energy loss of the penetrating muons, obtained through 4-fold gate condition, has been shown in Fig. \ref{UGSpecExp}~(a). Real time of data collection was 1128 hours (47 days). The features of the spectrum include characteristic bump at low partial energy deposit ($\Delta E$) at the scintillator, followed by a bump at larger $\Delta E$, which has the characteristics of a Landau distribution, similar to the feature exhibited in the measurements done at the Earth's surface (see Fig.~\ref{SurfOp}(b)).

To obtain the actual flux of cosmic muons at the 555 m deep underground laboratory, one has to devise a method to subtract the first bump at low $\Delta E$, having significant overlap with the Landau bump. Our approach was based on some kind of pattern recognition through the contribution of the decay radiation from known or standard radioactive sources to the energy deposition spectrum of the 4-fold scintillator assembly. It is well known and also established from our measurements at the underground laboratory, shown in Section \ref{sec:Gamma}, that soft and discrete gamma rays from ${}^{40}\rm K$ and ${}^{238}\rm U$, ${}^{232}\rm Th$ decay chain nuclei are present. Therefore, gamma reference source based measurements were undertaken to find out whether the cosmic muon telescope, operated in 4-fold coincidence mode, would be sensitive to these gamma rays. Measurements were done using a ${}^{22}\rm Na$ gamma reference source placed both on top of the scintillator stack and also in between the two pairs of  scintillators (middle of the whole scintillator assembly) and the detectors were operated keeping the electronic configurations same as before. Based on these measurements, we could generate the $\Delta E$ spectrum acquired with a 4-fold gate due to the source alone. The spectrum is shown in the right panel of Fig. \ref{UGSpecExp}. It is apparent that the spectral shape, acquired over a time span of a few tens of hours, has discernible similarity with the first bump of the spectrum in Fig. \ref{UGSpecExp} (a). Spectral distribution of Fig. \ref{UGSpecExp} (b) was fitted with a pulse fit function which is a combination of exponential growth followed by exponential decay. The parameter values, their uncertainties and reduced $\chi$-squared were found to be reasonable. Following this, the low $\Delta E$ bump region of Fig.~\ref{UGSpecExp} (a) was fitted with the same function and best fit parameters were obtained. 

The functional curve over the spectral region of the first bump in Fig.~\ref{UGSpecExp} (a), estimated using the best fit parameters, was subtracted from the overall spectral distribution itself to obtain the exclusive $\Delta E$ spectral distribution. The spectrum, shown by the green line in Fig.~\ref{UGSpecExp} (a), closely resembles in shape with $\Delta E$ spectral distribution of cosmic muons recorded at the surface (see Fig.~\ref{SurfOp}(b)). The measured cosmic muon flux at the underground laboratory, obtained by integrating this background subtracted $\Delta E$ spectrum, was calculated to be $(2.257\pm0.261\pm0.042)\times10^{-7}\hspace{1mm} \rm cm^{-2}\hspace{0.5mm}sec^{-1}$. The first of the two errors indicates the systematic uncertainty estimated from the variation in the background subtraction arising due to uncertainties in the best fit parameter values, while the second error represents the statistical uncertainty.

\subsection{Monte Carlo Simulation}
\label{sec:MuonSimul}
Cosmic muon flux measurement, as described in Sec.~\ref{expt}, essentially provides the integrated intensity of the cosmic muons reaching the detector over its solid angle of acceptance. It fails to yield any information about the energy spectrum of the penetrating particles. Fortunately, there are well-developed tracking simulation like GEANT4 toolkit\cite{Geant4Pap} 
to carry out realistic Monte Carlo simulation of muon propagation through the rock overburden. A reliable event generator, rock composition and the average density of the rock ($2.89 \pm 0.09~{\rm gm.cm}^{-3}$) are required. While the last two input specific to the site have already been reported\cite{JUSL_Collab}, the event generator for the purpose was constructed following the flux distribution of muons at the sea level, as given by Gaisser's formula\cite{PDG_Cosmic}:
\begin{equation}
	\frac{d^2N_{\mu}}{dE_{\mu,0}d\Omega}=\frac{0.14E^{-2.7}_{\mu,0}}{{\rm cm^2\hspace{0.5mm}sec\hspace{0.5mm}sr\hspace{0.5mm}GeV}}\times\Bigg\{\frac{1}{1+\frac{1.1\hspace{0.5mm}E_{\mu,0}\cos\theta}{\epsilon_{\pi}}}+ \frac{\eta}{1+\frac{1.1\hspace{0.5mm}E_{\mu,0}\cos\theta}{\epsilon_{\mathcal{K}}}}\Bigg\}\,\,,
\label{EqGeis}
\end{equation} 
where $E_{\mu,0}$ represents the energy of muons at sea level $\epsilon_{\pi}=115$ GeV, $\epsilon_{\mathcal{K}}=850$ GeV and $\eta=0.054$. The two terms within bracket in the above equation account for the muon production by the charged pions and kaons respectively during passage through the atmosphere. This formula contains zenith angle ($\theta$) dependence and approximates
\begin{figure}[ht]
	\centering
	\includegraphics[scale=0.5]{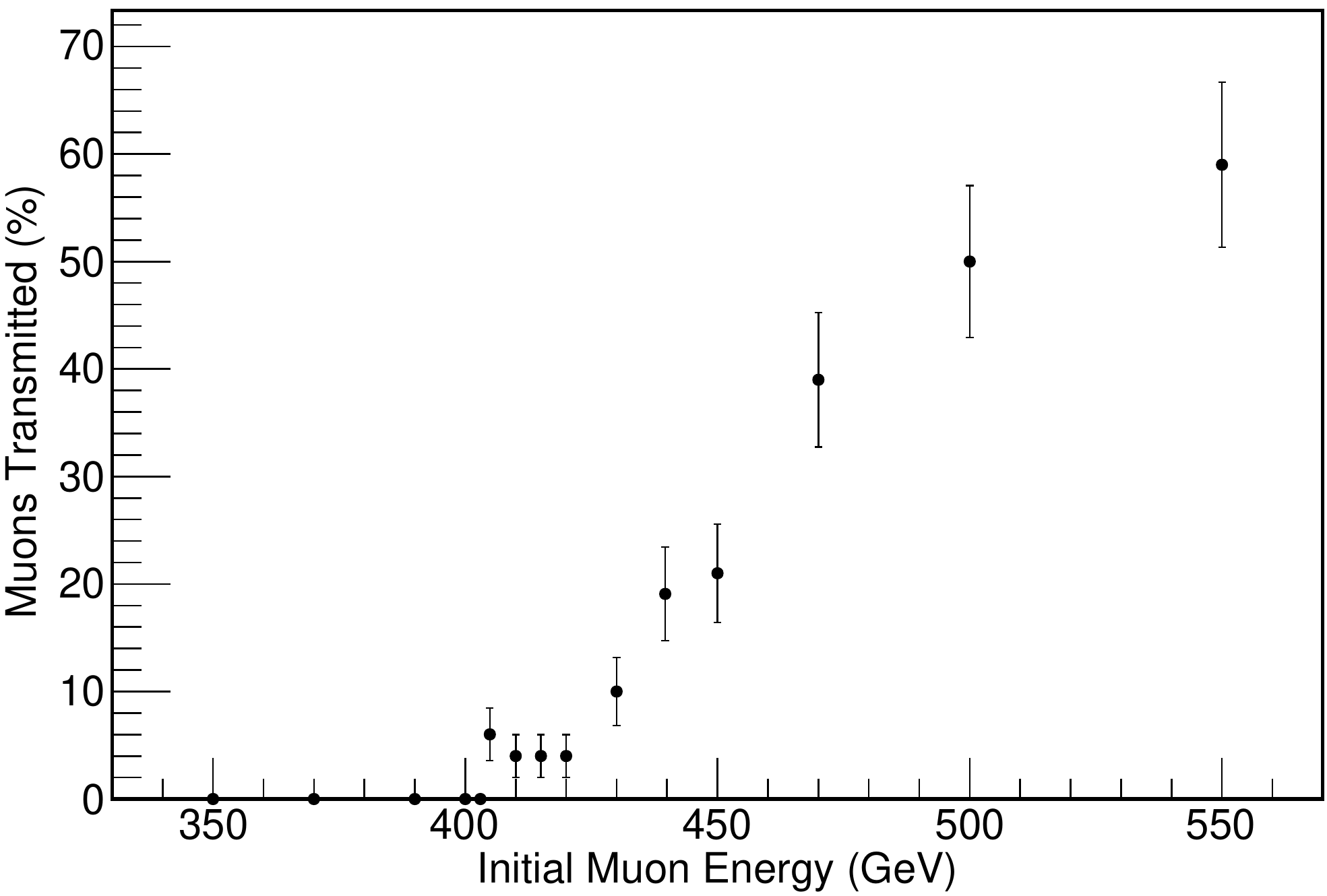}
	\caption{Percentage of mono-energetic muons transmitted through to 555 m underground level.}
	\label{TransPlot}
\end{figure}
the sea level muon flux for energies $E_{\mu} >$ $\frac{100\hspace{0.5mm} {\rm GeV}}{\cos\theta}$ and for $\theta\hspace{0.5 mm}<\hspace{0.5mm}70^{\circ}$. These muons would interact with the rock material and begin to lose energy. Many of the low energy muons would be stopped inside the rock before reaching the experimental hall located at the 555 m deep underground. For a rough estimate of the minimum energy a muon must have at sea level to reach the experimental hall, we have propagated streams of mono-energetic muons, with different energies, vertically through 555 m of rock, having the same composition as shown in Table 1 of
\begin{figure}[ht]
\centering
\includegraphics[scale=0.5]{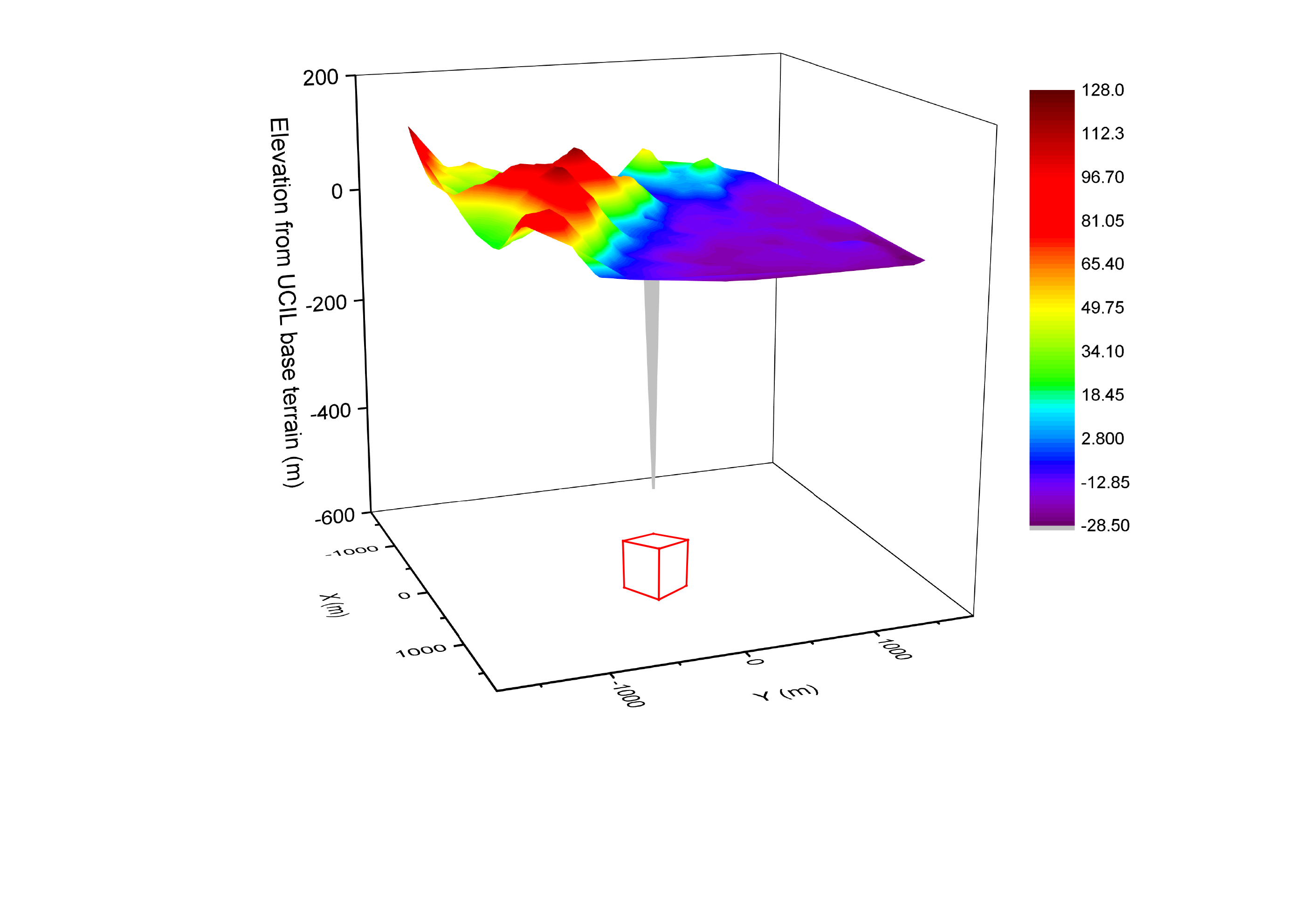}
\vspace{-10 mm}
\caption{The terrain of area above the underground site for the  laboratory. Relative depth of the laboratory is indicated.}
\label{Terrain}
\end{figure}
Ref.~\cite{JUSL_Collab}. Fraction of these muons of different energies, which would penetrate through 555 m of rock, is plotted in Fig.~\ref{TransPlot}. It can be seen from the figure that only those muons with $E_{\mu,0} \gtrsim 400$ GeV could penetrate the entire rock overburden and reach the site at JUSL. Although the zenith angle distribution for muons at sea level is given by $I(\theta)=I(0)\cos^2\theta$, the zenith angle distribution varies almost as $I(\theta)=I(0)\sec\theta$~\cite{PDG_Cosmic} at such high energies ($E_{\mu,0}\gg\epsilon_{\pi}$). The total integrated flux at sea level in the energy range $E_{\mu,0}>400$ GeV , as estimated from Eq.~\ref{EqGeis}, is roughly around $3.03\times10^{-6}\hspace{1mm} \rm cm^{-2}\hspace{0.5mm}sec^{-1}$.

\begin{figure}[ht]
	\centering
	\includegraphics[scale=0.4]{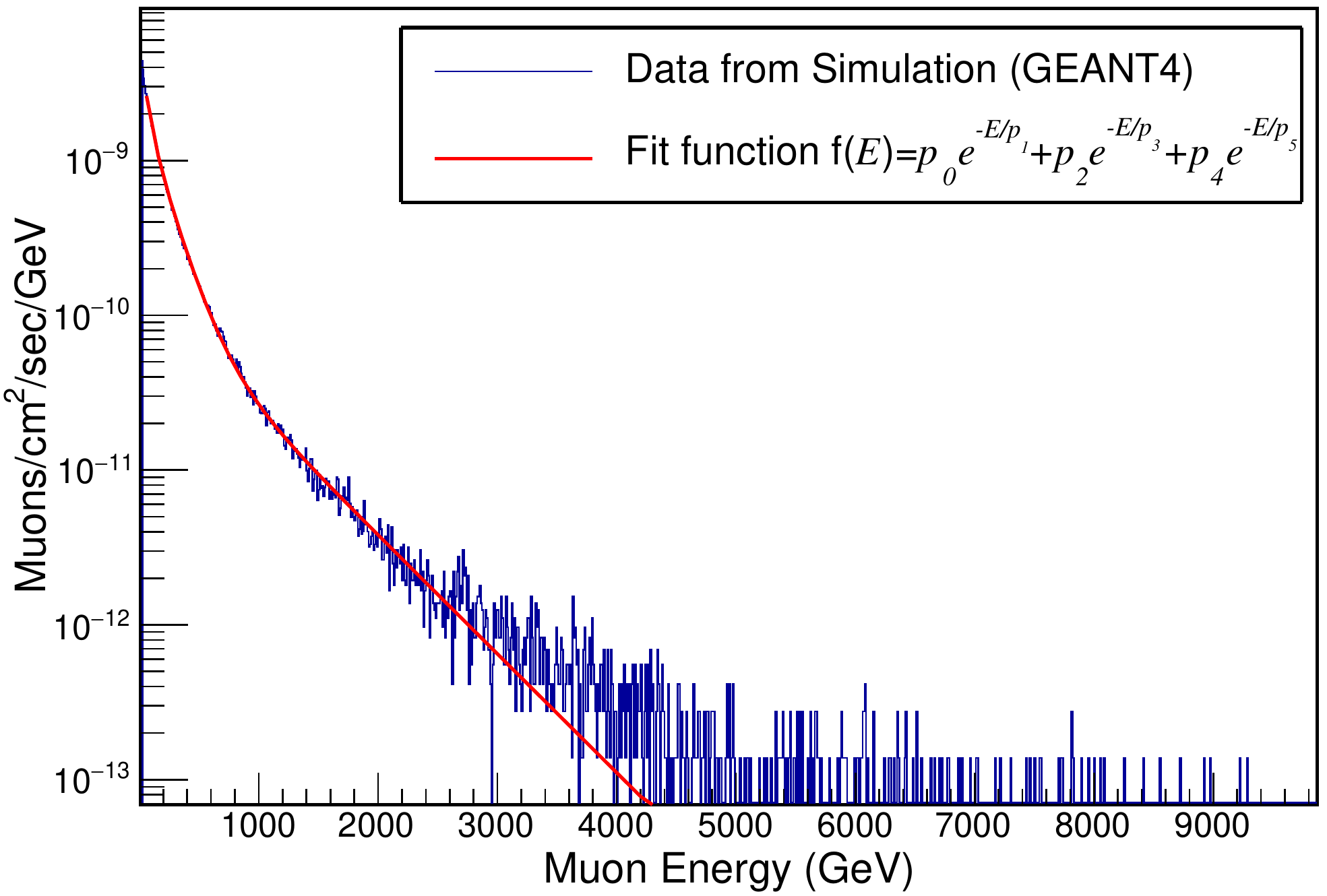}
	\hspace{2 mm}
	\includegraphics[scale=0.4]{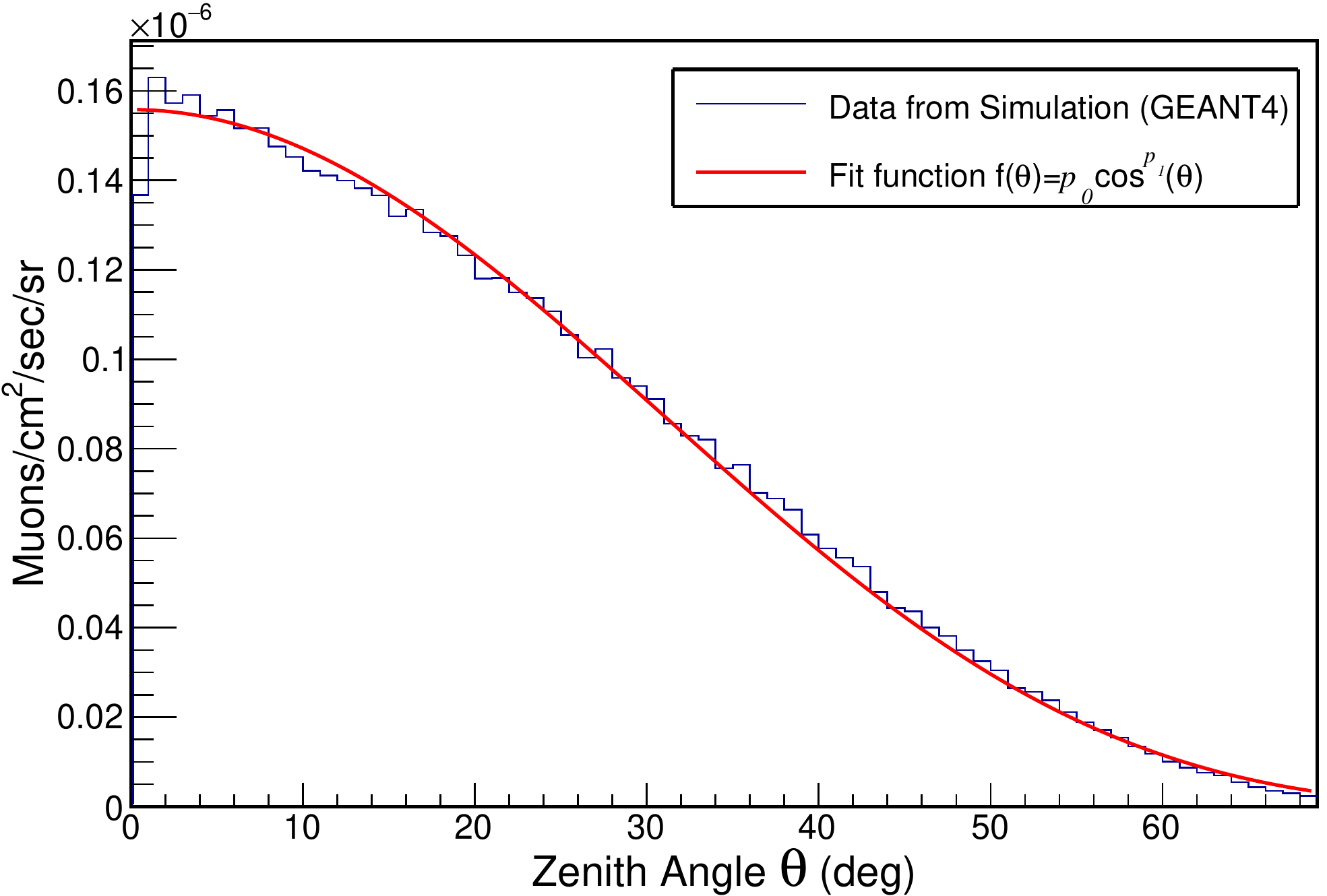}
	\caption{The cosmic muon energy (left panel) and zenith angle (right panel) distributions at the underground site. The energy spectrum is fitted with a function as shown in the figure, while the function used to fit the zenith angle distribution is $f(\theta)={\rm A}\cos^n(\theta)$.}
	\label{UGPlots}
\end{figure}

Determining the energy and the zenith angle distribution of cosmic muons at the experimental site at JUSL involves simulating the passage of cosmic muons through the rock. Energy and zenith angle ($\theta$) distributions of the cosmic muons at the surface are used to construct the event generator for our simulation. Rock composition as enlisted in Table~1 of Ref.~\cite{JUSL_Collab}, was used. In addition, the terrain information of the surrounding area, as mapped from Google Earth Pro~\cite{GEarth}, was used. This is shown in the 3D map of Fig.~\ref{Terrain}.  In our simulation, we have generated muons in the range $400\hspace{1mm}{\rm GeV}<E_{\mu,0}<15\hspace{1mm}{\rm TeV}$ in accordance with Gaisser's formula on top of a 3 km $\times$ 3 km surface. During the passage through the rock material, the muons would interact, lose energy and also produce secondary particles. Only those events where the muons were able to penetrate through the rock material and emerged out on the other side, were recorded. The energy and the zenith angle distribution at the underground site, as obtained from simulation, is shown in Fig. \ref{UGPlots}. The energy spectrum is fitted with sum of three exponential functions as shown in the left panel of Fig. \ref{UGPlots}, and the average energy of muons at 555 m depth was found to be around $186.45\pm0.51$~GeV. The quoted error is due to the variation in rock composition. The zenith angle distribution at the underground site was found to vary as: $I(\theta)\simeq I(0)\cos^n\theta$, with $n$ = $3.756\pm0.047\pm0.009$. The total integrated flux of cosmic muons at the underground laboratory was found to be: $(5.927\pm0.376\pm0.009)\times10^{-7}$ $\rm cm^{-2}\hspace{0.5mm}sec^{-1}$ with a vertical intensity of $(1.558\pm0.099\pm0.002)\times10^{-7}$ $\rm cm^{-2}\hspace{0.5mm}sec^{-1}\hspace{0.5mm}sr^{-1}$. The errors shown indicate the systematic and the statistical uncertainties respectively. The systematic uncertainty arises from the uncertainties in the determination of the rock composition by various techniques, discussed in
Ref.~\cite{JUSL_Collab}, and variation in density of the rock. In addition, the gamma ray background, arising from interaction of the muons with the rock material during their propagation, was also accounted for separately. The gamma ray flux inside the laboratory  was estimated as: $(2.59\pm0.16)\times10^{-6}$ $\rm cm^{-2}\hspace{0.5mm}sec^{-1}$ with an average energy of $\sim$8~MeV. This flux is almost 2 orders of magnitude lower than the measured gamma ray background flux at the underground site for $E_\gamma \gtrsim 3\,{\rm MeV}$ (see Sec.~\ref{sec:Gamma}).

The discrepancy between the simulated and the experimental muon flux is primarily due to the geometric effect of the experimental detector setup which was not taken into account so far. The 4-fold scintillation method misses a significant percentage of muon events since the solid angle coverage is less than 2$\pi$ and also the asymmetric nature of the scintillator itself. We have performed a numerical simulation to estimate the solid angle coverage leading to the aperture effect of the detector. A schematic 
representation of the detector and the different co-ordinate variables used for the numerical estimation are shown in Fig. \ref{DetScheme}. The variables $a$ and $b$ represent the length and width respectively of each scintillator while $c$ is the total height of the assembly. In addition, $\theta_{\rm max}$ is the maximum zenith angle that a muon can have to register a 4-fold scintillation signal and $\phi_{\rm min}$ and $\phi_{\rm max}$ denote the corresponding minimum and maximum azimuthal angle. Needless to say that this representation only caters to one of the four coordinates and therefore the final
result needs to be scaled appropriately to get the real estimate of muon flux.  

It is to be noted that only a reduced area of the scintillator telescope will be useful in registering a four-fold coincidence signal for the muons and the azimuthal angle coverage $\Delta \phi = \phi_{\rm max} - \phi_{\rm min}$ is not same for all the opening angles of the telescope. In fact, the coverage reduces as $\theta$ increases. Taking all of these factors into account, coupled with the area normalization for inclined muons, the estimated muon flux as observed by the detector assembly can be written as
\begin{equation}\label{MuFluxInt}
\Phi^{\rm detector}_{\mu} = \frac{4\hspace{0.5mm}I_0}{a\times b}\times\int_{0}^{\theta_{\rm max}}\int_{\phi_{\rm min}}^{\phi_{\rm max}}(a-c\tan\theta\cos\phi)(b-c\tan\theta\sin\phi)\cos^{n+1}\theta\sin\theta\hspace{0.5mm}d\theta\hspace{0.5mm}d\phi\,\,.
\end{equation}
In the above equation the reduction in the length and width inside the integral accounts for the reduced area as explained and $n$ stands for the exponent in the zenith angle distribution of muons at the 555 m deep underground lab, found by fitting the simulation data as presented in Fig.~\ref{UGPlots}~(right panel). The extra $\cos\theta$ contribution is a result of area normalization for muons at an inclination $\theta$. 
We have also performed a Monte Carlo calculation to determine the aperture function of Eq.~\ref{MuFluxInt} using the energy and the zenith angle distributions from the GEANT4 simulation (see Fig.~\ref{UGPlots}). Excellent agreement was found between the two estimates to extract the vertical muon intensity $I_0$.   
\begin{figure}[ht]
	\centering
	\includegraphics[scale=0.5]{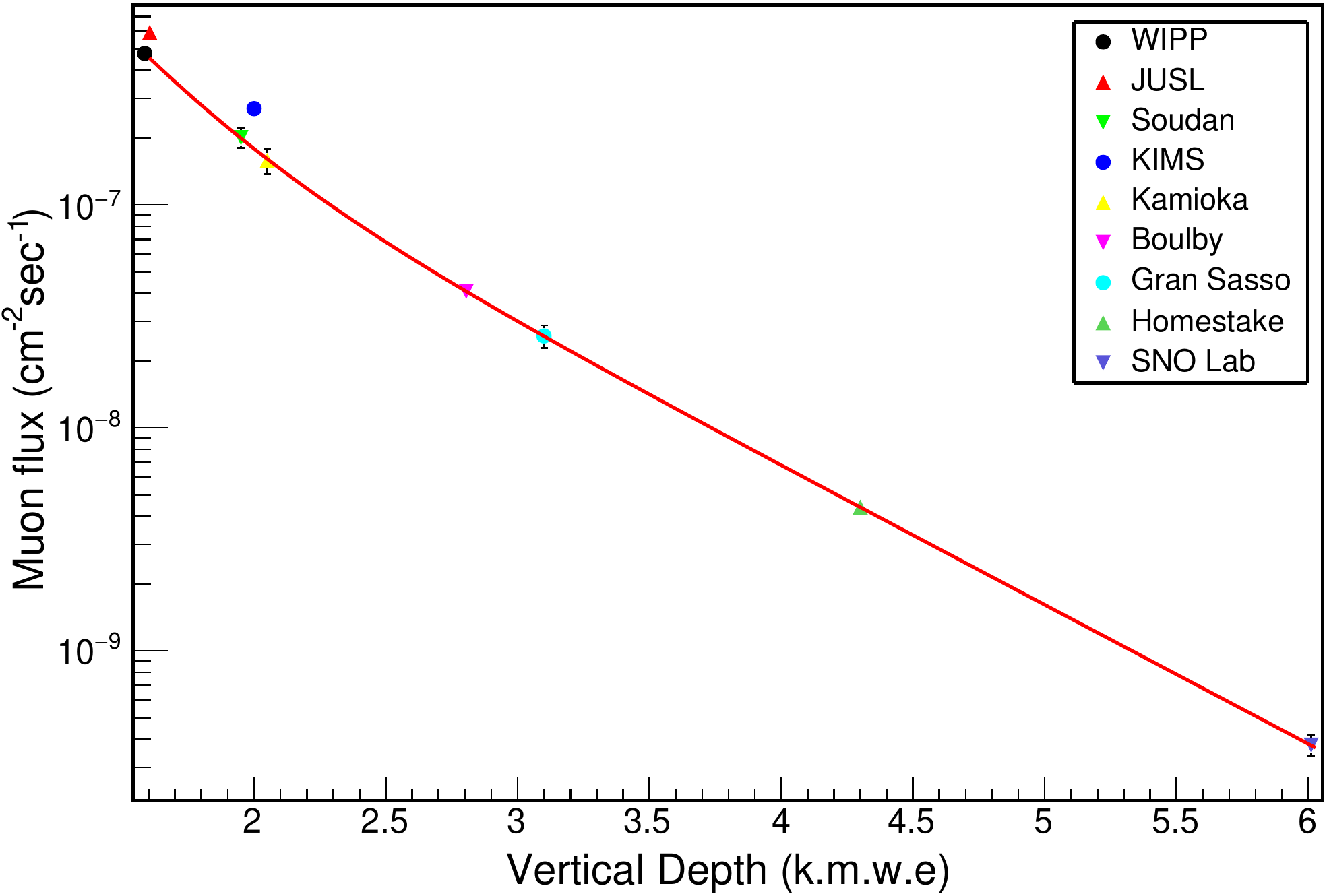}
	\caption{Total muon flux at different underground laboratories around the world are shown in the plot. Results are from WIPP\cite{WIPP_Muon},Soudan mine\cite{ Soudan_Muon}, KIMS\cite{KIMS_Limits}, Kamioka mine\cite{ Kamioka_firstRes}, Boulby\cite{Boulby_Muons}, Gran Sasso\cite{Mei_Hime}, Homestake\cite{Mei_Hime}, and SNOLab\cite{SNO_Res} with their respective vertical depths in kilometer water equivalent(kmwe).The global fit function is given in Eq. \ref{GlobMuFit}. Our result is also shown for comparison.}
	\label{DiffLabMu}
\end{figure}
Therefore, substituting $I_0 = (1.558\pm0.099\pm0.002)\times10^{-7}$ $\rm cm^{-2}\hspace{0.5mm}sec^{-1}\hspace{0.5mm}sr^{-1}$, found from the GEANT4 simulation into Eq. \ref{MuFluxInt}, the muon flux is obtained as $(2.051\pm0.142\pm0.009)\times10^{-7}$ $\rm cm^{-2}\hspace{0.5mm}sec^{-1}$. It can be seen that the experimentally determined muon flux: $(2.257 \pm 0.261 \pm 0.042) \times 10^{-7}\,{\rm cm^{-2}.\, sec^{-1}}$ is in good agreement with the GEANT4 simulation results within the respective range of uncertainties.

Equipped with the cosmic muon flux results obtained for the JUSL site at 555 m depth, we have included our result for comparison with similar results of cosmic muon intensity at different vertical depths $h$ (expressed in km of water equivalent), as plotted in the vertical depth $(h)$ vs. intensity $(I_\mu)$ curve shown in Fig.~\ref{DiffLabMu}. The global fit function used can be expressed as~\cite{Mei_Hime}
\begin{equation}\label{GlobMuFit}
I_{\mu}(h)=a_1e^{-h/b_1}+a_2e^{-h/b_2}\,\,,    
\end{equation}
where $h$ is the vertical depth in kmwe, considering a flat rock overburden and the global fit parameters $a_1$, $a_2$, $b_1$ and $b_2$ are adopted from Ref.~\cite{Boulby_Muons}. Reasonable agreement of our result with those from other such underground laboratories placed at different geographical locations, confirms our measurements and related simulations. Furthermore, the measured cosmic muon flux and the estimated energy spectra obtained from our simulation, provide further input for simulation and estimation of flux density of the cosmogenic neutrons (see Sec.~\ref{sec:CosmoNeut}).

\section{Neutron Background and its measurements}
\label{neutflux}
The neutron flux in the approximate range of $\lesssim 10 \, {\rm MeV}$ energy was measured using pressurized $^4$He detector manufactured by Arktis Radiation Detectors, Switzerland\cite{Chan}. The active element of the detector consists of a stainless steel cylindrical cell, 600 mm long and of 65 mm inner diameter filled with $^4$He gas at 150 - 180 bar pressure. Fast neutrons entering the detector volume undergo elastic scattering from the $^4$He nuclei, resulting in nuclear recoil within the pressurized gas cell.
The recoiling and energetic $^4$He nuclei deposit energy into the medium by ionization and excitation to singlet and triplet states (excimers), resulting in production of scintillation light in the VUV region $(\lambda \sim 80 \, {\rm nm})$\cite{Kell}. The inner walls of the cell was coated with Wavelength shifting (WLS) materials to convert the scintillation light to wavelengths acceptable to the array of silicon photomultipliers used as photon readout.  

The singlet and the triplet excimer states decay with two different time constants resulting in a fast and a slow component. Furthermore, the ratio of population of the two excimer states differ for $\gamma-$rays or electrons and the neutrons, which was exploited to achieve $n-\gamma$ discrimination. In addition, one of the major advantage of using a light atom with very few electrons, such as $^4$He, is that the detector is expected to be much less sensitive to $\gamma-$rays than the neutrons of comparable energy\cite{Kell2}. Consequently, the detector achieves very good $n-\gamma$ discrimination capability. 

However, efficiency of the detector for neutrons has energy dependence as reflected from the energy dependence of the elastic scattering cross section, which peaks around 1~MeV and falls off at increasing energy\cite{Chan}. Estimated average efficiency over the energy spectrum of neutrons emitted from a $^{252}$Cf spontaneous fission source, was found to be $\sim10$\%. In addition, the detector has the capability of detecting thermal neutrons due to the addition of a Lithium compound to the material used for internal coating\cite{arktis}. Energetic charged particles, produced by $^6{\rm Li} (n, ^3{\rm H}) ^4{\rm He}$ reactions having almost 4 orders of magnitude larger cross section for the thermal neutrons than that for the fast neutrons, cause scintillation in the gas medium, thereby increasing the relative presence of slow components in the signal. On board digital pulse shape discrimination provide the time over threshold (ToT) logic signal for the corresponding pulse above the preset electronic threshold. As mentioned above, predominance of fast component for the incident $\gamma-$rays result in short duration of the ToT signal, relatively larger duration of ToT is expected for the fast neutrons because of abundance of the slow component in the signal, and even larger duration of ToT for the thermal neutrons because of larger thermal neutron cross 
\begin{figure}[ht]
	\centering
	\includegraphics[scale=0.7]{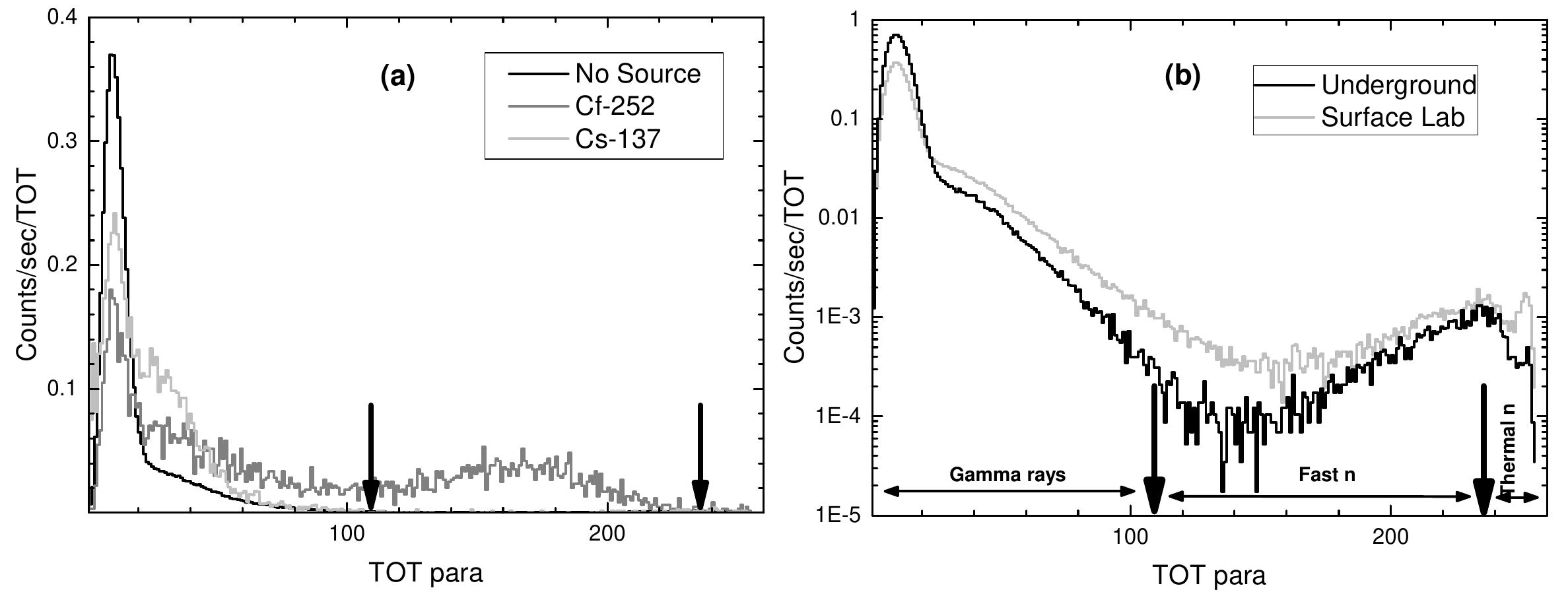}
	\caption{Spectral plots of the Time over Threshold (ToT) signals for (a) $^{137}$Cs $\gamma-$reference source and $^{252}$Cf spontaneous fission source, and (b) neutron background at the underground cavern. The ToT spectral plot obtained at the overground laboratory is shown for comparison.}
	\label{nspcref}
\end{figure}
sections which varies as $1/v$, $v$ being the velocity of the neutrons. Typical spectral distributions of the ToT signals for different radioactive reference sources, such as $^{137}$Cs $\gamma-$ray source and $^{252}$Cf neutron source, are shown in the Fig.~\ref{nspcref}(a). Discrimination of $\gamma-$rays, fast neutrons and thermal neutrons by the ToT signal markers are indicated. The gamma response is clearly manifested for the low ToT signal range. The spectral distribution for $^{252}$Cf fission source, which  emits both $\gamma-$rays and fast neutrons are evident from the plot. The marked zones of ToT signals for $\gamma-$rays, fast neutrons and thermal neutrons are indicated in the plot. 

ToT spectral distribution measurements were carried out at the underground laboratory and the laboratory over ground at the same location. The plots are presented in the Fig.~\ref{nspcref}(b). Since there is significant overlap of the ToT signals between the zones of response for the $\gamma-$rays and the fast neutrons, an exponential fit to the tail of the $\gamma-$response zone was done and its contribution overlapping with the ToT signal zone for the fast neutrons was subtracted from the integral count to determine the fast neutron flux, both at the underground and the over-ground laboratory. Measured fast neutron flux at the underground laboratory was $(9.93 \pm 0.22 \pm 0.10) \times 10^{-5}\, {\rm cm}^{-2} \, {\rm sec}^{-1}$. The first quoted uncertainty is the statistical error and the second one is the systematic error arising from the subtraction procedure at the overlap region as mentioned above. The energy threshold cut for fast neutrons could not be estimated as the neutron energy and related calibration procedure could not be done. However, the threshold for fast neutrons is estimated over the broad range of $\sim 0.1 - 1\, {\rm MeV}$ corresponding to the situation where there is overlap between the $\gamma-$response and the fast neutron-response zones (see Fig.~\ref{nspcref}~(b)). 
On the other hand, thermal neutron flux measured at the site using the same detector was $(6.15 \pm 0.18) \times 10^{-5}\, {\rm cm}^{-2} \, {\rm sec}^{-1}$, and therefore, the total flux of radiogenic neutrons as measured at the underground site was: $(1.61 \pm 0.03) \times 10^{-4}\, {\rm cm}^{-2} \, {\rm sec}^{-1}$ corresponding to no threshold cut. 

\section{Simulation of residual neutron flux}
\label{neutfluxsim}
As mentioned in Sec.~\ref{intro}, the residual neutron background at the underground laboratory, caused by the intrinsic rock radioactivity is termed as the radiogenic neutron background, while the neutrons produced by the interaction of the penetrating cosmic muons with the rock materials is called the cosmogenic neutron background. The fluxes and the energy spectra of the cosmogenic and the radiogenic neutron background depend on a) the depth of the site and the rock composition, and b) concentration of ${\rm U}/{\rm Th}$ and the rock composition respectively.      

\subsection{Simulation for radiogenic neutrons}\label{sec:RadNeut}
\begin{figure}[ht]
	\centering
	\includegraphics[scale=0.55]{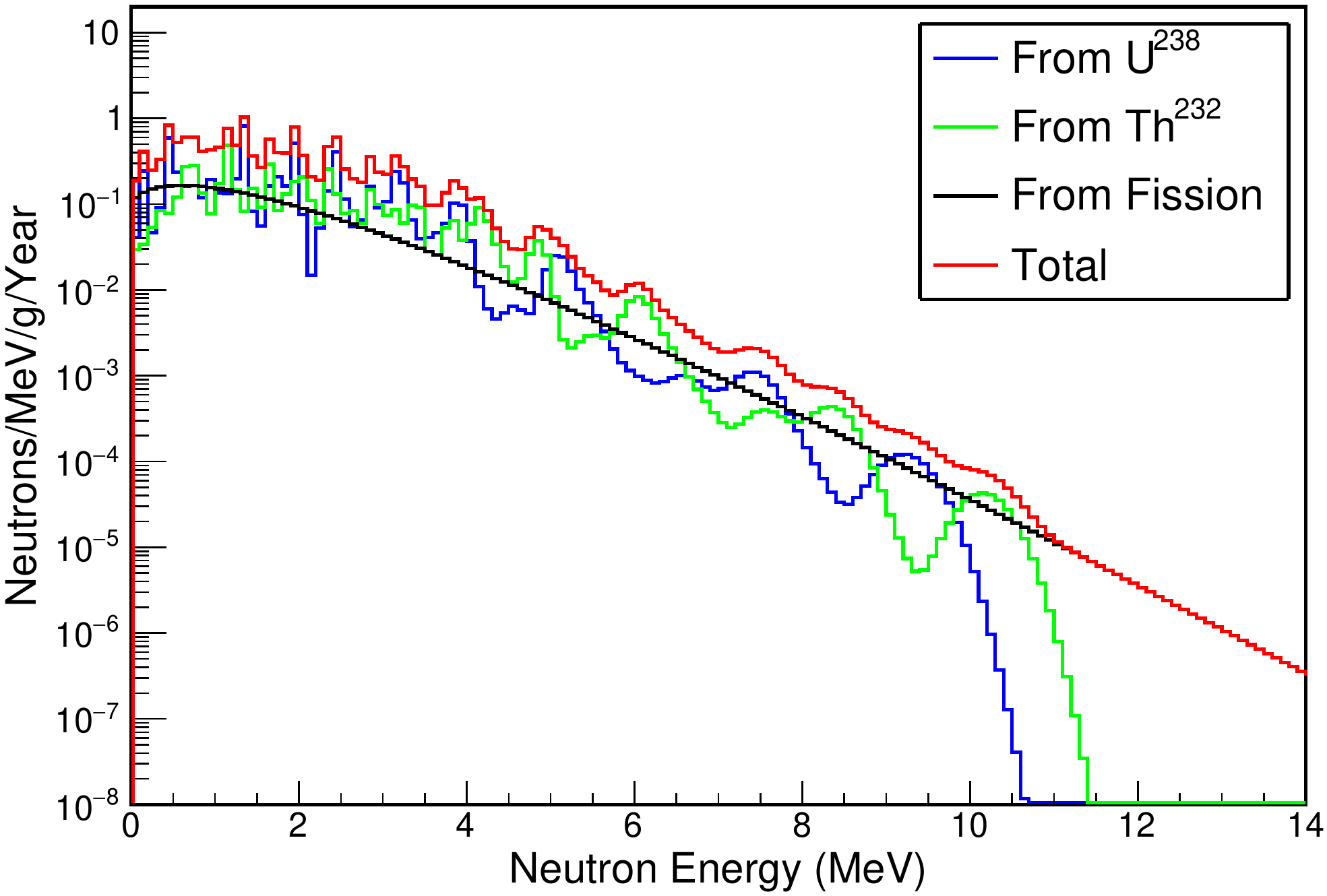}
	\caption{Energy spectrum of neutrons produced due to the ($\alpha,n$) reactions and spontaneous fission of the rock components~\cite{JUSL_Collab}.}
	\label{NeuProdRock}
\end{figure}

Neutrons are generated by the interaction of the energetic $\alpha-$particles produced from decay of the remnant $^{238}$U and $^{232}$Th ($\alpha,n$) reactions with the nuclei present in the rock constituents and also spontaneous fission of naturally occurring $^{238}\rm U$, $^{232}\rm Th$ and their daughters in the respective decay chains. Simulation of radiogenic neutron generation from the rocks surrounding the laboratory site require detailed knowledge of the composition and the quantitative estimates of presence of the radioactive elements in the surrounding rocks. Details of these aspects can be found in our earlier work\cite{JUSL_Collab}. The energy spectrum of neutrons produced in the rock due to the above processes are given in  Fig. \ref{NeuProdRock}. 
\begin{figure}[ht]
	\centering
	\includegraphics[scale=0.6]{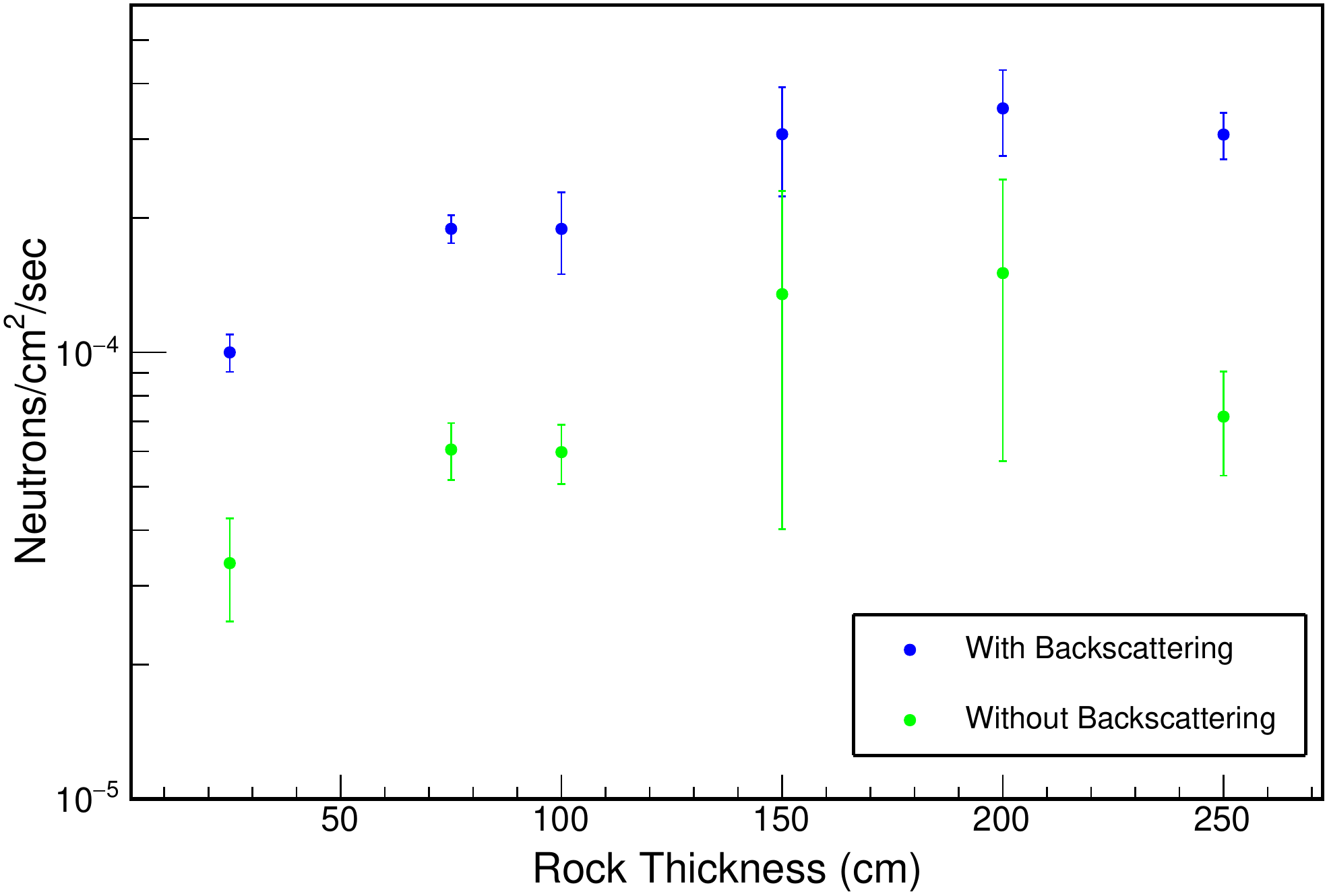}
	\caption{Fraction of radiogenic neutrons crossing the walls and the floor of the cavern as function of the shell thickness of rock considered in the tracking simulation.}
	\label{shellopt}
\end{figure}

It is evident that the simulation process requires propagation of the neutrons generated from the points of production to the interior of the laboratory cavern and eventually we need to estimate the number of neutrons falling on a detector. We have done GEANT4 tracking simulation for this purpose. A hemispherical cavern, of 2.2 m radius with shells of surrounding rocks of varied thicknesses, was considered. Individual radioactive sources of estimated concentration (see Table 1 of Ref.~\cite{JUSL_Collab}) were embedded randomly in the surrounding shell of rock including the floor.   
Inside the cavern, that is inside the laboratory volume, a cylindrical neutron detector of 70 mm diameter and 600 mm length was placed. This resembles the detector used for neutron detection. Neutrons, generated at random points distributed homogeneously inside the shell of rock surrounding the laboratory volume, were isotropically propagated. Kinetic energies of the neutrons were randomly sampled from the spectral distribution shown in Fig. \ref{NeuProdRock}. 
\begin{table}
	\centering
	\scalebox{1.1}{
		\renewcommand{\arraystretch}{1.2}
		\begin{tabular}{|c|c|c|C{1.5cm}|c|C{1.5cm}|}
			\hline
			\multirow{4}{1.5cm}{\bf Category} & \multirow{4}{1.1cm}{\hspace{2.5mm}$\rm \bf E_{th}$ \\ (MeV)} & \multicolumn{2}{c|}{\textbf{Laboratory}} & \multicolumn{2}{c|}{\textbf{Detector}}\\
			\cline{3-6}
			& & $\bf\Phi_n$ ($\rm cm^{-2}\hspace{0.5mm}sec^{-1}$) & Mean Energy (MeV) & $\bf \Phi_n$ ($\rm cm^{-2}\hspace{0.5mm}sec^{-1}$) & Mean Energy (MeV)\\
			\hline
			\multirow{3}{1.1cm}{\bf \hspace{4.5mm}I} & 0.0 & $(1.50\pm0.93)\times10^{-4}$ & 0.184 & $(2.56\pm0.16)\times10^{-4}$ & 0.165 \\
			& 0.1 & $(4.91\pm1.35)\times10^{-5}$ & 0.913 & $(4.79\pm0.48)\times10^{-5}$ & 0.851 \\
			& 1.0 & $(1.57\pm0.31)\times10^{-5}$ & 2.032 & $(1.41\pm0.15)\times10^{-5}$ & 1.971  \\  \hline
			\multirow{3}{1.1cm}{\bf \hspace{3.5mm}II} & 0.0 & $(3.51\pm0.76)\times10^{-4}$ & 0.140 & $(2.61\pm0.17)\times10^{-4}$ & 0.162 \\ 
			& 0.1 & $(7.58\pm1.01)\times10^{-5}$ & 0.806 & $(4.82\pm0.49)\times10^{-5}$ & 0.848 \\ 
			& 1.0 & $(2.05\pm0.35)\times10^{-5}$ & 1.967 & $(1.41\pm0.14)\times10^{-5}$ & 1.969  \\ \hline			
	\end{tabular}}
	\caption{The radiogenic neutron fluxes ($\bf\Phi_n$) and their average energies within the laboratory cavern and as registered by the cylindrical detector for different neutron energy thresholds ($\rm \bf E_{th}$). The categories {\bf I} and {\bf II} classify the flux estimates without and with backscattering events respectively. See text for details.}
	\label{NeuFluxTab}
\end{table}
The neutrons, propagating through the shell of rock, undergo elastic or inelastic scattering multiple times before entering the laboratory volume. GEANT4 tracking was done to record the fate of each event after the scattering. While a small fraction of the neutrons stop well inside the shell of rock, rest of the neutrons enter the laboratory volume. Fraction of stopped neutrons vary with the thickness of the shell. Number of neutrons finally entering the laboratory volume through the inner wall and the floor of the cavern shows a trend towards saturation as the thickness of the shell of rock is increased. (see Fig.~\ref{shellopt}). 
Based on this result, we have considered the thickness of the shell of rock as 2 m for all our simulation work for the radiogenic neutrons. 

The flux of radiogenic neutrons were estimated in different ways and summarized in Table~\ref{NeuFluxTab}. Neutrons passing through the walls and floor of the cavern per unit area per second to enter into the laboratory volume was estimated along with the spectral distribution. The estimated flux inside the laboratory, obtained through integration of the spectral distribution for different energy thresholds, and the average energy are listed in the Table under category I. Since the entering neutrons would have random directions of entry, a simple scaling by the area of the neutron detector will not work and therefore, further tracking of the neutrons inside the cavern was done. Majority of the incoming neutrons within the cavern would suffer multiple scattering from the walls.  Only fraction of the surviving neutrons impinging on the detector surfaces was included in the flux estimates of neutrons under category I, as seen by the detector for different energy thresholds. 
 
However, the estimates of laboratory flux under category I do not include backscattering of neutrons within the cavern. These are the neutrons which enter the laboratory volume through the wall, and after passing through it, hit another wall or the floor to get re-scattered. The same process was found to repeat until these neutrons get absorbed.  Similarly, a neutron which impinge on the detector inside the cavern, may get scattered off after registering a count in the detector. The process may repeat multiple times till the neutron gets absorbed. These events result in multiple hits from the same neutron that entered into the cavern and the cylindrical detector volume. These neutrons are tagged separately as {\em back-scattered} neutrons in our tracking simulation. Finally, the radiogenic neutron fluxes inside the laboratory and falling on the detector volume were estimated by including these back-scattered neutrons. These fluxes, considered as category II, were presented separately in Table~\ref{NeuFluxTab}.
\begin{figure}[ht]
	\centering
	\includegraphics[scale=0.6]{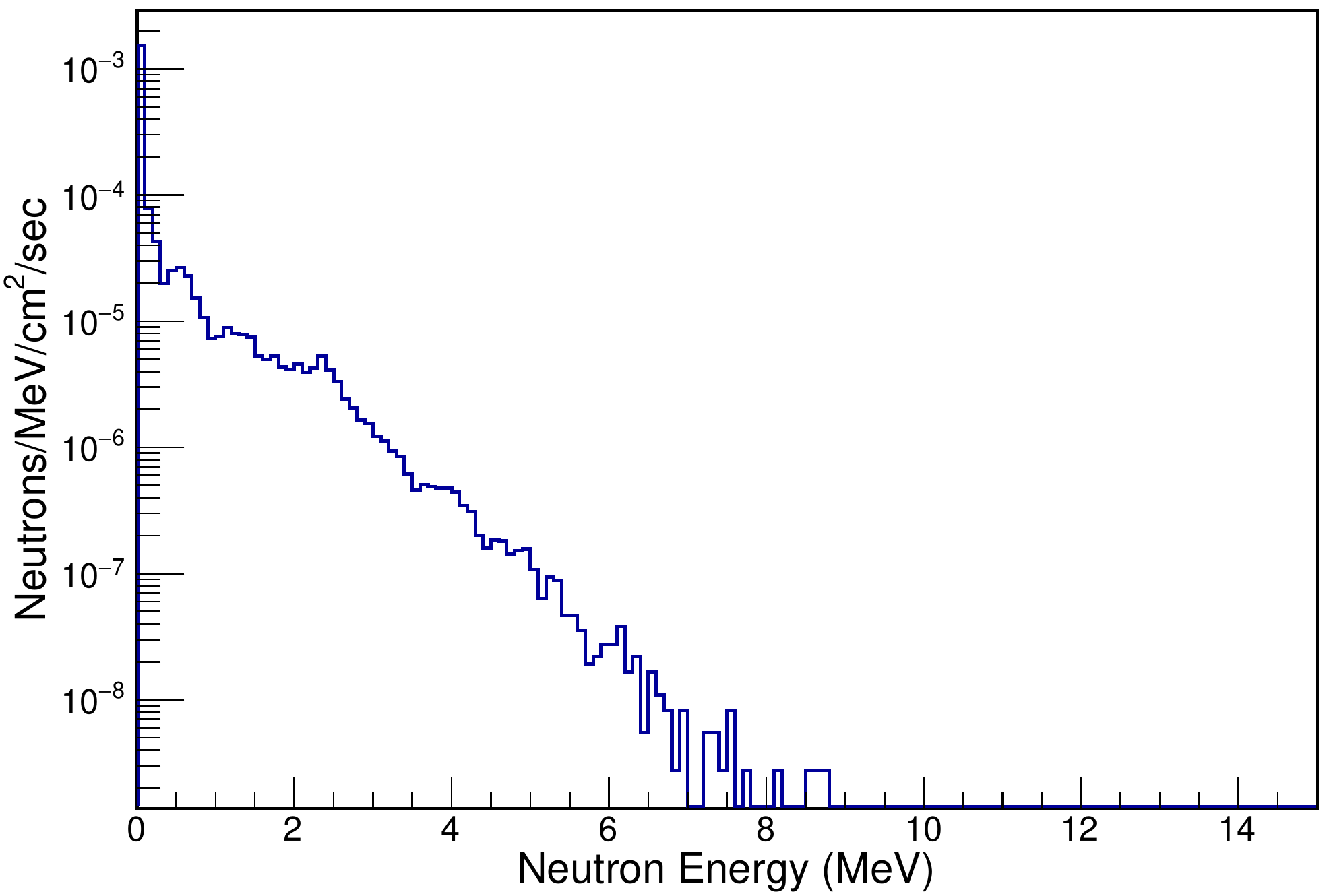}
	\caption{Simulated energy spectrum of radiogenic neutrons falling on the detector placed inside the laboratory.}
	\label{NeuDecSpec}
\end{figure}


From the flux estimates presented, it can be seen that major contribution to the flux without any threshold cut comes from the back-scattered neutrons, which dominate the spectral distribution of radiogenic neutrons at sub-MeV energy. The average energy of the neutrons also reveal the same fact. This is very much in agreement with the expectation that multiple scattering will result in slowing down of the neutrons. From our tracking simulation, the estimated energy spectrum of the neutrons falling on the detector inside the cavern is shown in Fig. \ref{NeuDecSpec}.
Furthermore, the radiogenic neutron flux including backscattered neutrons (category II) with no threshold cut is $\sim 2.3$~times the corresponding flux obtained by excluding the backscattered neutrons (category I). For threshold cuts at 0.1 and 1 MeV, these same factors are $\sim 1.5$ and $\sim 1.3$ respectively which indicate that the backscatter fraction gradually decreases with increase in the energy of the radiogenic neutrons. Similar conclusions were also reported in earlier studies at Boulby Laboratory in the UK\cite{Boulby_Neut} and also at the Kamioka Observatory\cite{kamioka}. Therefore, even at a raised detection threshold of 1 MeV, almost 30\% of the residual radiogenic neutron flux inside the cavern is due to the backscattered neutrons. However, the flux values even at zero threshold tend towards some agreement, if we consider the fluxes due to the neutrons falling on the detector. The backscattering contributes an additional $\sim ~2$\% to the flux at zero threshold. This is well within the quoted error in the estimates. 
   
For comparison with experimentally measured neutron flux, which is: $(1.61 \pm 0.03) \times 10^{-4}$~neutrons.${\rm cm}^{-2}.{\rm sec}^{-1}$ corresponding to zero threshold cut, we consider the radiogenic neutron flux estimated by our simulation, as seen by the detector with no threshold cut (see Table~\ref{NeuFluxTab}).  Our simulation yields the flux as: $(2.61 \pm 0.17) \times 10^{-4}$~neutrons.${\rm cm}^{-2}.{\rm sec}^{-1}$.  
The result of simulation agrees reasonably well with the experimentally measured neutron flux within the limitations of both measurements and simulation. Though we consider the neutron flux measured by the detector as that corresponding to zero threshold, in practice, the flux due to the fast neutrons were measured with certain uncalibrated threshold cut corresponding to a) reduction of electronic noise, and b) ToT signal overlap of neutrons with gamma rays which could not be avoided. This would reduce the measured fast neutron flux, while the thermal neutron flux measurements do not possibly suffer from such an issue.  

\subsection{Simulation for cosmogenic neutrons}\label{sec:CosmoNeut}

As mentioned in the introduction (Sec.~\ref{intro}), the cosmogenic neutrons are produced mostly by the penetrating muons interacting with the rock. Therefore, it is natural to expect dependence of the cosmogenic neutron flux on the penetration depth. Cosmogenic neutrons are mostly produced by the four processes: 1) muon capture leading to ejection of neutrons by pre-equilibrium reactions or fission, 2) muon spallation reactions, 3) neutron production through muon or related photon induced hadron cascades, and 4) neutron produced by the muon induced photons through electromagnetic cascades\cite{Kudra}. These interaction inputs are quite accurately included in the recent versions of GEANT4\cite{Geant4Pap} or FLUKA\cite{fluka}, which have been used to estimate the cosmogenic neutron spectra at major underground laboratories\cite{Kamioka_firstRes,Kudra,Lux,SNOn}. These results are used extensively to estimate the cosmogenic neutron flux due to the surrounding rock and also that produced inside the detectors, mostly the large volume detectors (LVD). A systematic and comprehensive simulation study of the cosmic muon flux and consequent cosmogenic neutron background at various underground laboratories at depths up to 6 km.w.e were done\cite{Mei_Hime}, which predicts a simple scaling behavior of the cosmic muon flux as a function of vertical depth (see Fig.~\ref{DiffLabMu}).
Majority of these studies rely on the Gaisser formula with the universal scaling behavior as mentioned above to generate the muons underground for simulation. We have used somewhat mixed approach, where the integrated muon flux from our measurements (see Sec.~\ref{muons}) is used, but the energy distribution of the muons entering the laboratory is obtained from GEANT4 simulation, with muons at the surface being generated following Gaisser's formula. The measured muon flux was found to agree reasonably well with GEANT4 simulation as shown in Sec.~\ref{sec:MuonSimul}.

Strategy for estimating the flux and the spectral distribution of cosmogenic neutrons require some optimization. This is because of the fact that tracking propagation of the cosmic muons through huge rock overburden spanning a few hundreds of metres would require a lot of computing power, if not computing time to achieve the desired results. Instead of propagating the cosmic muons and tracking the secondary neutrons all along the 555 m vertical depth, we have considered a shell of rock of varying thickness surrounding the cavern and allowed the muons to propagate through the shell to produce hadronic showers. Our simulation takes into account the spectral distribution of cosmic muons reaching the outer boundary layer of the shell of thickness, varying in the range of a few tens of centimetres to 4 metres. Average energy of the entering muons is around $180$~GeV and therefore, the energy of the cosmogenic neutrons is expected to be approximately in the range of a few MeV to a few GeV.

We have traced the neutrons produced in the process inside the shell, leading towards entry inside the cavern. The neutrons that are produced inside the rocks interact with the rock material and lose energy. These neutrons may get absorbed inside the shell of rock or the interactions may lead to the production of secondary neutrons. Finally, those neutrons which cross the inner boundary to reach the interior of the cavern are counted to estimate the neutron flux inside the laboratory. Depending on the solid angle of acceptance, a fraction of the flux inside the laboratory would be counted by the detector. 

\begin{figure}[ht]
	\centering
	\includegraphics[scale=0.4]{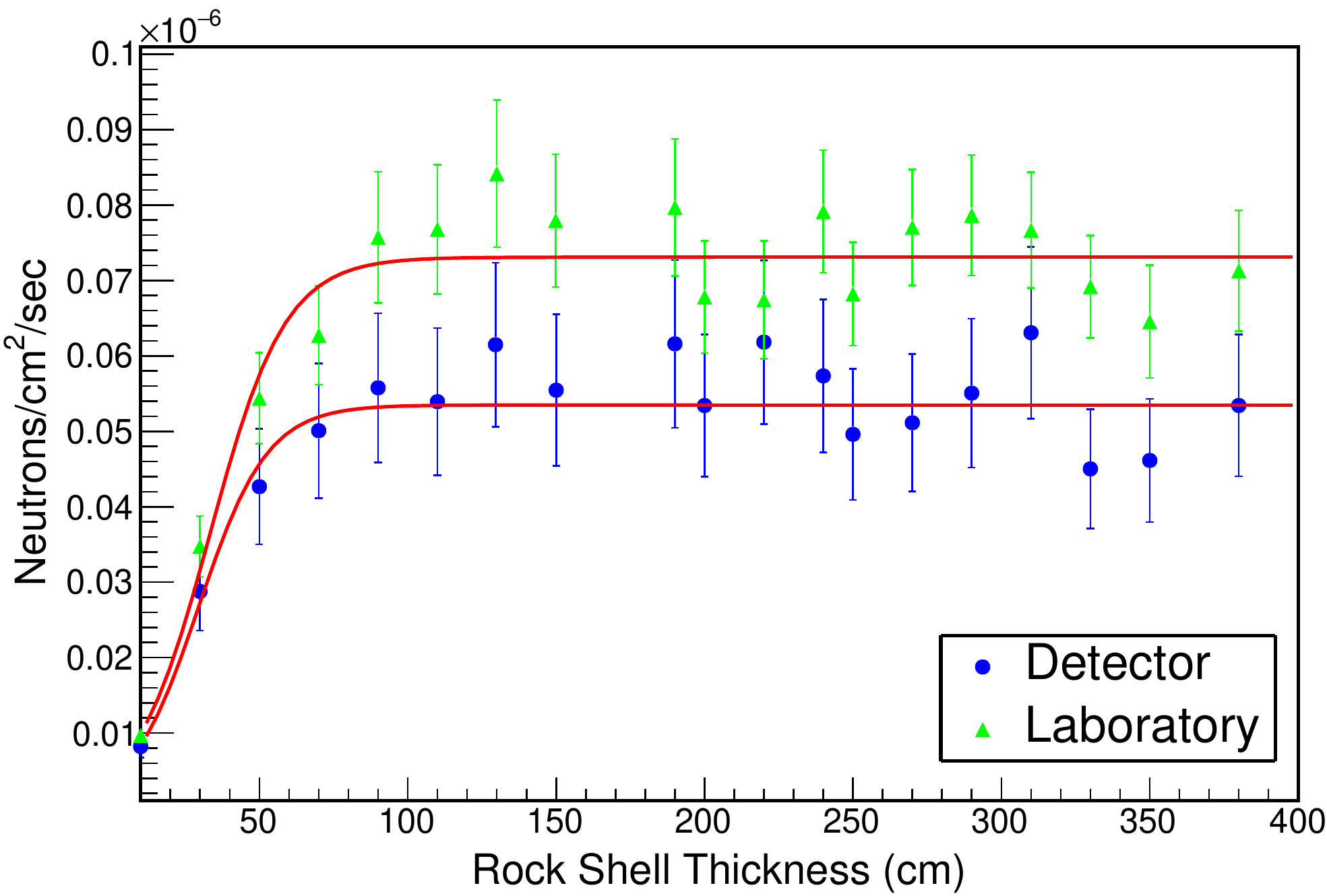}
	\hspace{2mm}
	\includegraphics[scale=0.4]{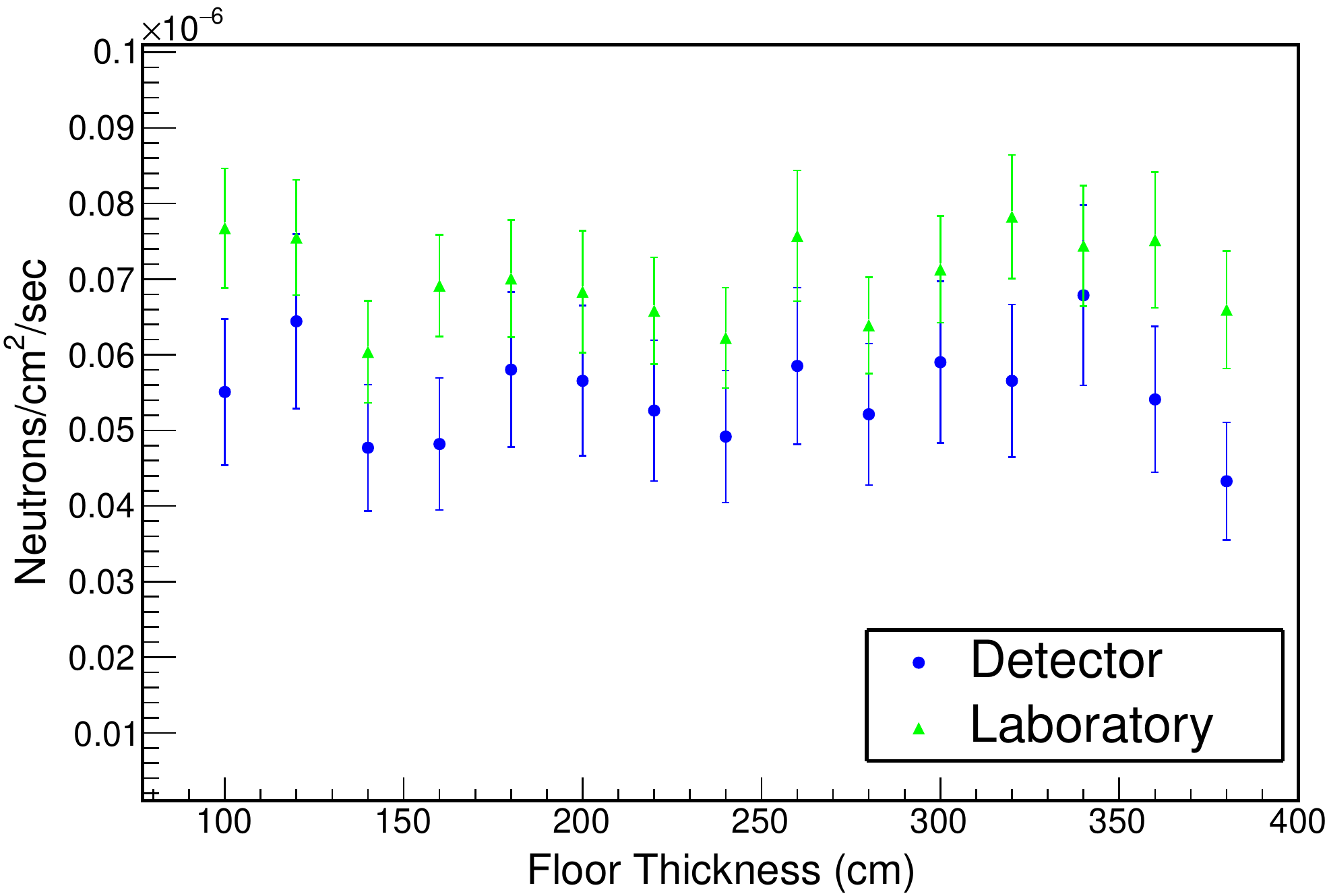}
	\caption{Cosmogenic neutron flux as a function of rock shell thickness (Left panel) and floor thickness keeping the rock shell thickness fixed at 2 m (Right panel).}
	\label{CosmoNeutThick}
\end{figure}
To begin with, we fix the floor thickness at 2 metres, and estimate the cosmogenic neutron fluxes for varying thicknesses of the shell. Results are shown in the left panel of Fig.~\ref{CosmoNeutThick}. The cosmogenic neutron fluxes at the laboratory and at the detector inside the laboratory show a saturating behaviour starting around 100~cm thickness. We have considered the optimized hemispherical shell thickness of 200~cm for the simulation. Variation of the flux inside the cavern and as seen by the detector for variable thicknesses of the floor, keeping the shell thickness fixed at 2 metres, is shown in the right panel of Fig.~\ref{CosmoNeutThick}. It is evident from the plot that the flux remains mostly independent of the floor thickness, which is qualitatively expected because the corresponding neutrons are backscattered from the floor. This results in relatively softer energy spectrum of these neutrons and therefore, causing saturation at relatively smaller thicknesses.

\begin{table}
	\centering
	\scalebox{1.1}{
		\renewcommand{\arraystretch}{1.2}
		\begin{tabular}{|c|c|c|C{1.5cm}|c|C{1.5cm}|}
			\hline
			\multirow{4}{1.5cm}{\bf Category} & \multirow{4}{1.1cm}{\hspace{2.5mm}$\rm \bf E_{th}$ \\ (MeV)} & \multicolumn{2}{c|}{\textbf{Laboratory}} & \multicolumn{2}{c|}{\textbf{Detector}}\\
			\cline{3-6}
			& & $\bf\Phi_n$ ($\rm cm^{-2}\hspace{0.5mm}sec^{-1}$) & Mean Energy (MeV) & $\bf \Phi_n$ ($\rm cm^{-2}\hspace{0.5mm}sec^{-1}$) & Mean Energy (MeV)\\
			\hline
			\multirow{3}{1.1cm}{\bf \hspace{4.5mm}I} & 0.0 & $(3.07\pm0.87)\times10^{-8}$ & 22.067 & $(5.48\pm0.06)\times10^{-8}$ & 3.887 \\ 
			& 0.1 & $(1.09\pm0.19)\times10^{-8}$ & 44.249 & $(1.19\pm0.08)\times10^{-8}$ & 17.851 \\ 
			& 1.0 & $(7.57\pm0.45)\times10^{-9}$ & 63.800 & $(5.80\pm0.26)\times10^{-9}$ & 36.239  \\ \hline
			\multirow{3}{1.1cm}{\bf \hspace{3.5mm}II} & 0.0 & $(8.45\pm0.83)\times10^{-8}$ & 6.456 & $(5.66\pm0.07)\times10^{-8}$ & 3.766 \\ 
			& 0.1 & $(1.83\pm0.18)\times10^{-8}$ & 26.759 & $(1.21\pm0.08)\times10^{-8}$ & 17.677 \\ 
			& 1.0 & $(9.48\pm0.42)\times10^{-9}$ & 51.524 & $(5.83\pm0.27)\times10^{-9}$ & 36.097  \\ \hline
	\end{tabular}}
	\caption{The cosmogenic neutron fluxes ($\bf\Phi_n$) and their average energies in the laboratory space and as detected by the cylindrical detector for different detection thresholds ($\rm \bf E_{th}$). The categories {\bf I} and {\bf II} classify the flux estimates without and with backscattering events respectively. See text for details.}
	\label{CosFluxTab}
\end{table}

\begin{figure}[ht]
	\centering
	\includegraphics[scale=0.6]{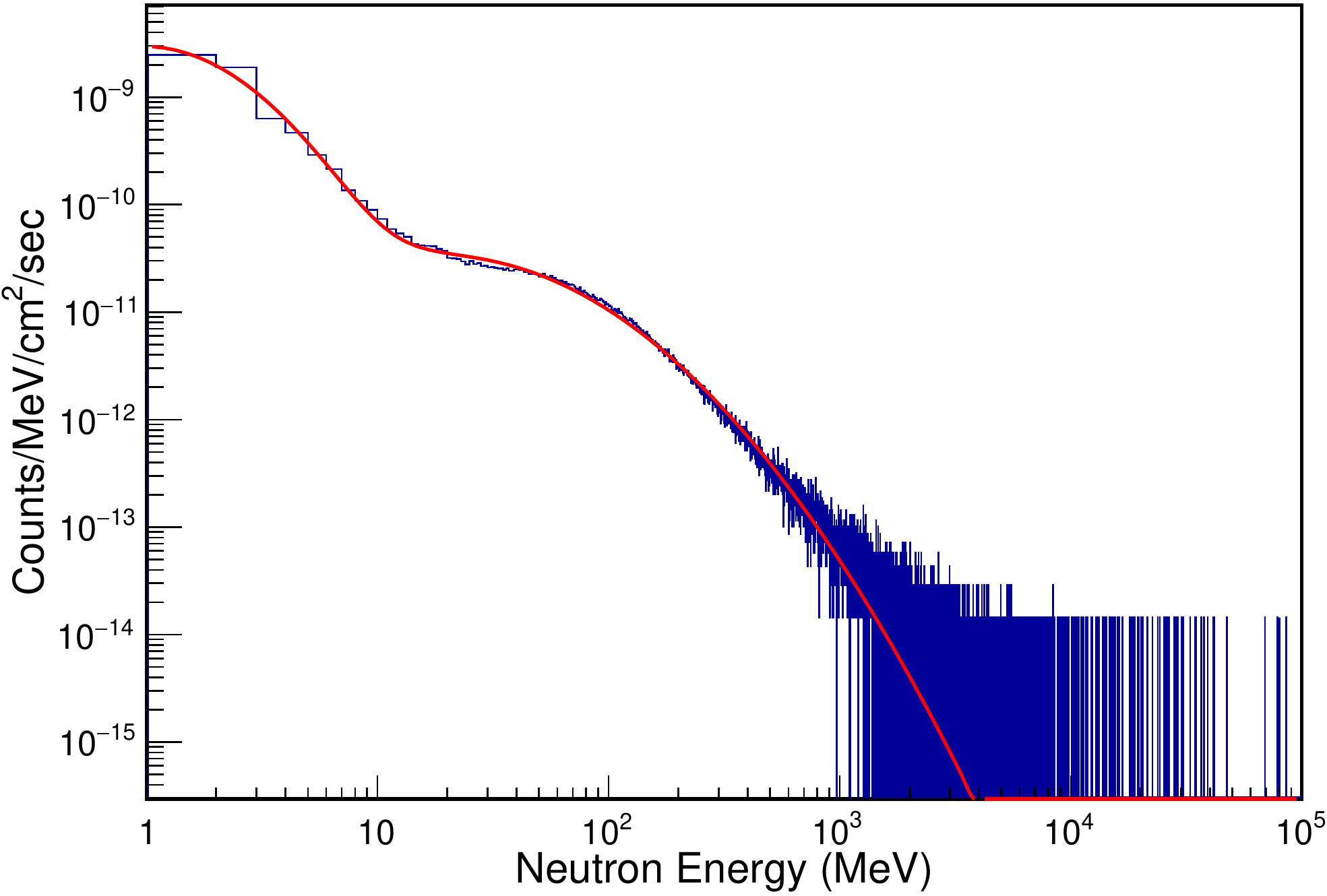}
	\caption{Energy spectrum of cosmogenic neutron in the laboratory volume. The fit function (red) is the same as that given in~\cite{Gordon_CosNeut_ana,Heim_2014PRD}.}
	\label{CosmoNeutSpec}
\end{figure}

\begin{figure}[ht]
	\centering
	\includegraphics[scale=0.6]{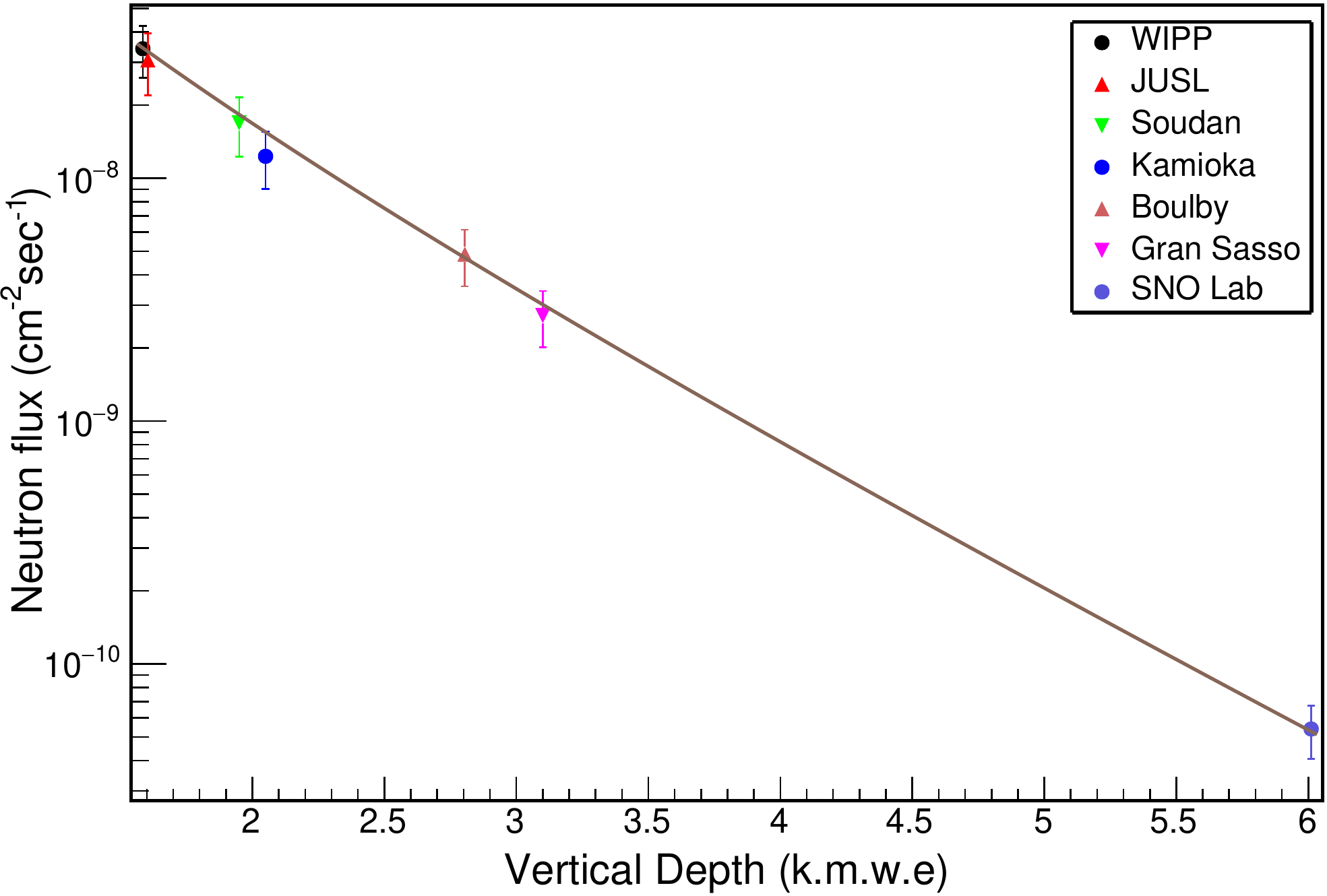}
	\caption{Comparison of the cosmic-muon induced neutron background at JUSL with the simulation results for other underground sites.~\cite{Mei_Hime}. Multiple hits due to the back-scattering effect has not been accounted for here as in~\cite{Mei_Hime}.}
	\label{DiffLabCosmoNeutFlux}
\end{figure}
A comparison of the cosmogenic neutron fluxes inside the laboratory and those seen by the detector at different energy thresholds are presented in Table~\ref{CosFluxTab}. 
\begin{table}[ht]
    \centering
    \begin{tabular}{|c|c|c|c|}
    \hline
         $j$ & $c_j$ & $\beta_j$ & $\gamma_j$ \\
         \hline
         \hline
         1 & $(2.964\pm0.038)\times10^{-9}$ & $(0.785\pm0.021)$ & $(-0.047\pm0.036)$\\
         2 & $(9.418\pm2.099)\times10^{-13}$ & $(0.413\pm0.011)$ & $(2.425\pm0.099)$\\
    \hline
    \end{tabular}
    \caption{Parameters obtained from fitting the cosmic muon-induced neutron energy spectrum of Fig. \ref{CosmoNeutSpec} with the analytical model in Eq. \ref{NeuSpec_ana}}
    \label{CosNeutSpecFitPar}
\end{table}
The fluxes obtained with and without accounting for back-scattering are presented. The total cosmogenic neutron flux at the laboratory and at the detector placed inside the laboratory, after accounting for back-scattering events, are $(8.458\pm0.826\pm0.003)\times10^{-8}$ $\rm cm^{-2}\hspace{0.5mm}sec^{-1}$ and $(5.661\pm0.068\pm0.035)\times10^{-8}$ $\rm cm^{-2}\hspace{0.5mm}sec^{-1}$ respectively for zero threshold, $(1.839\pm0.185\pm0.077)\times10^{-8}$ $\rm cm^{-2}\hspace{0.5mm}sec^{-1}$ and $(1.204\pm0.081\pm0.016)\times10^{-8}$ $\rm cm^{-2}\hspace{0.5mm}sec^{-1}$ at 0.1 MeV threshold respectively. The uncertainties indicated represent the systematic and the statistical uncertainties respectively. The estimated energy spectrum of the cosmic muon induced neutrons inside the laboratory volume is shown in Fig. \ref{CosmoNeutSpec}. The spectrum has been fitted with an analytical model~\cite{Gordon_CosNeut_ana,Heim_2014PRD}, which can be expressed as 
\begin{equation}\label{NeuSpec_ana}
\frac{d\Phi_n}{dE_n}=\sum^{2}_{j=1}c_j \times{\rm exp}\left[-\beta_j({\rm ln}(E))^2+\gamma_j{\rm ln}(E)\right]\,\,,
\end{equation}
where the parameters $c_j$, $\beta_j$ and $\gamma_j$, obtained from the fit, are enlisted in Table~\ref{CosNeutSpecFitPar}.     Unlike the treatment presented in Ref.~\cite{Heim_2014PRD}, all of the parameters in Eq. \ref{NeuSpec_ana} were derived from the fit. 
These parameters, though similar in orders of magnitude and signs, differ from those values given in the Ref.~\cite{Heim_2014PRD} possibly due to the fact that both the vertical depths and the rock compositions are different between the two sites. The estimated errors on the parameters are obtained from optimization of the fitting procedure. However, it may be noted that the normalization factors $c_j$ extracted from our analysis match in orders of magnitude with those listed in Ref.~\cite{Heim_2014PRD}. 

From the estimated cosmogenic neutron fluxes given in Table~\ref{CosFluxTab}, it can be seen that back scattering effect increases the flux seen by the detector at zero threshold by $\sim 4\%$, whereas for 0.1 and 1 MeV threshold, the fluxes are the same within the respective limits of uncertainty.
A comparison of the above cosmogenic neutron flux results with that of the simulation results for other underground sites is shown in Fig.~\ref{DiffLabCosmoNeutFlux}. The global fit function used is of the form~\cite{Mei_Hime}
\begin{equation}
\Phi_n(h)=p_0\left(\frac{p_1}{h}\right)e^{-h/p_1}\,\,,
\end{equation}
where $h$ is the vertical depth in kmwe considering flat overburden, $p_0$ and $p_1$ are the global fit parameters. 
For the comparison, cosmogenic neutron flux inside the respective laboratories were estimated by neglecting the effect of multiple hits due to backscattering of neutrons from the cavern walls. Estimated neutron flux inside the JUSL cavern for zero energy threshold agree reasonably well with the global fit\cite{Mei_Hime} and therefore, validates our simulation procedure. However, as in the case of radiogenic neutrons, we rely on and recommend using the flux values estimated including backscattering effects for comparison with any experimental result and design strategies for rare event search experiments proposed to be set up at the laboratory in the future.

\section{Conclusion}
\label{concl}
We have reported a comprehensive evaluation of the radiation background at a newly established underground site in India at 1.6 kmwe vertical depth, with a prototype laboratory designated for setting up rare event search experiments. The study includes experiments involving measurements of in-situ gamma ray flux, cosmic muon flux, and the radiogenic neutron flux. Detailed Monte Carlo simulation for the cosmic muons, radiogenic neutrons produced from the residual U-Th decay chains, and the energetic cosmogenic neutrons produced by the penetrating cosmic muons have been carried out with the necessary details.

The gamma ray flux measurements include studies of spectral distributions of the discrete gamma rays produced from the U-Th decay chain remnants and the primordial nuclei, which extend up to 2.6 MeV. It is demonstrated from GEANT4 simulation that use of passive shielding involving Lead of reasonable thickness would reduce the gamma ray flux to $\lesssim 10^{-6}\,{\rm counts.\,cm^{-2}.\,sec^{-1}}$. The flux of gamma rays with $E_\gamma \gtrsim 3\,{\rm MeV}$, which are expected to be produced as secondaries from the cosmic muon background alone, is expected to be $\sim 10^{-8}\,{\rm counts.\,cm^{-2}.\,sec^{-1}}$ inside the same Lead shield. Corresponding estimated gamma ray flux from our simulation turns out to be lower by almost 2 orders of magnitude if we consider the muons alone as the source of high energy gamma ray background (see Sec.~\ref{sec:MuonSimul}). Other sources of high energy gamma rays, such as extensive air showers, GRBs etc. may be responsible for the enhanced flux. 

Good agreement within the limits of respective uncertainties are found between the experimentally measured cosmic muon flux at the underground laboratory using a muon telescope with 4 layers of plastic scintillators and our GEANT4 based simulations, including the solid angle correction due to acceptance of the telescope configuration. Our result has been found to be in reasonable agreement with the global fit to the flux data as function of vertical depth, available from seven major underground laboratories. The residual energy spectrum of the cosmic muons reaching the laboratory has been estimated from our simulation, which is used as source for estimation of the neutron background caused by the cosmic muons interacting with the surrounding rocks.

The neutron background caused by the radioactivity of trace elements embedded in the surrounding rocks has been carefully evaluated. These neutrons, termed as radiogenic neutrons, are found to have lower energy ($\lesssim 10\,{\rm MeV}$) as compared to the neutrons generated by the cosmic muons (cosmogenic neutrons), but the corresponding flux is several orders of magnitude larger than that of the cosmogenic neutrons. Flux of radiogenic neutrons was measured in-situ using a pressurized Helium-4 detector. Detailed Monte Carlo simulation using GEANT4 toolkit was done for the laboratory cavern using experimentally measured composition of the surrounding rocks. Fluxes inside the laboratory cavern and as seen by the detector were estimated at different neutron energy thresholds without and with backscattering of neutrons. It is demonstrated from our simulation that inclusion of backscattering of neutrons almost doubles the flux inside the laboratory for no threshold cut and for 1~MeV threshold, the backscatter contribution to the flux is $\sim 30\%$. Backscatter contribution appears to cause only minor modification $(\sim 2\%)$, if we consider the flux as seen by the detector for zero threshold. 
However, the measured radiogenic neutron flux is found to be almost 60\% less than the simulation result for zero threshold, but it lies between the corresponding numbers for $0-1\,{\rm MeV}$ threshold. It is noted that unspecified measurement threshold due to the discrimination procedure might be the reason for the mismatch between experiment and simulation.

Finally, the cosmogenic neutron background has been obtained through GEANT4 simulation using the cosmic muon spectral distribution estimated at the outer shell boundary to the laboratory cavern following some optimization procedure and after achieving good agreement between the measured cosmic muon flux and the estimate from simulation. As in case of radiogenic neutrons, the cosmogenic neutron fluxes for different threshold cuts are evaluated both inside the laboratory cavern and at the detector placed inside. In determining the neutron flux, the backscattered neutrons play more important role here than in case of the radiogenic neutrons because the spectral distribution of cosmogenic neutrons are a lot harder in comparison. This effect has been taken into account as neutron multiplicity effect in determining the corresponding flux at various underground laboratories\cite{Mei_Hime}. Using our simulation results, the average neutron multiplicity due to backscattering alone is found to be: $2.75 \pm 0.82$ inside the cavern. When compared with the global fit to the cosmogenic neutron flux results from six different underground laboratories, after neglecting the respective neutron mulitiplicity contributions\cite{Mei_Hime}, excellent agreement of our result
with the global fit is found which validates the simulation procedure followed in our work.

\section{Acknowledgements}
\label{ackn}
We express our sincere gratitude to the engineers, scientists and staff of the Uranium Corporation of India Limited (UCIL), Jaduguda, Jharkhand, India for their unconditional support towards establishing the laboratory at the mine facility. We are indebted to Ajit Kumar Mohanty, former Director, SINP for his constant encouragement, scientific and administrative support in laying the foundation of the underground laboratory and for carrying out scientific research to establish the feasibility of such a laboratory in India. We are grateful to the Crystal Technology Section of BARC, India for providing the portable gamma ray measurement station and the India based Neutrino Observatory (INO) group of TIFR for providing us with the large area plastic scintillators for the muon telescope. One of us (SS) would like to acknowledge financial support from the DAE Raja Ramanna Fellowship (DAE-RRF) scheme to carry out this work.        

\bibliography{sayan2bibfilev1}

\end{document}